\documentclass[11pt]{article}
\usepackage{mathrsfs}
\usepackage{amssymb}
\usepackage{amsmath}
\usepackage{ascmac}
\usepackage{amsthm}
\usepackage[dvipdfmx]{graphicx}
\usepackage{setspace}
\usepackage{color}
\usepackage{url}
\usepackage{booktabs}

\usepackage{comment}
\usepackage{bm}
\usepackage{titlesec}
\titleformat*{\section}{\large\bfseries}
\titleformat*{\subsection}{\it}

\newtheorem{thm}{Theorem}
\newtheorem{lem}{Lemma}

\newtheorem{algo}{Algorithm}

\def\ta{{\tau}}

\def\al{{\alpha}}
\def\be{{\beta}}
\def\ga{{\gamma}}

\def\ep{{\varepsilon}}

\def\th{{\theta}}

\def\ze{{\zeta}}
\def\bal{{\text{\boldmath $\alpha$}}}
\def\bbe{{\text{\boldmath $\beta$}}}

\def\bro{{\text{\boldmath $\rho$}}}

\def\bvth{{\text{\boldmath $\vartheta$}}}

\def\alt{{\tilde \al}}

\def\Ga{{\Gamma}}

\def\bPsi{{\text{\boldmath $\Psi$}}}

\def\p{{\text{\boldmath $p$}}}

\def\t{{\text{\boldmath $t$}}}

\def\w{{\text{\boldmath $w$}}}
\def\x{{\text{\boldmath $x$}}}
\def\y{{\text{\boldmath $y$}}}
\def\z{{\text{\boldmath $z$}}}

\def\I{{\text{\boldmath $I$}}}

\def\ft{{\tilde f}}

\def\tr{{\rm tr\,}}

\def\1r{{\rm (1)}}
\def\2r{{\rm (2)}}
\def\3r{{\rm (3)}}
\def\4r{{\rm (4)}}
\def\5r{{\rm (5)}}

\def\non{{\nonumber}}
\def\ep{{\varepsilon}}

\def\th{{\theta}}

\begin{document}
\title{On Data Augmentation for Models Involving Reciprocal Gamma Functions}
\author{
Yasuyuki Hamura\footnote{Corresponding author. 
Graduate School of Economics, Kyoto University, Yoshida-Honmachi, Sakyo-ku, Kyoto, 606-8501, JAPAN. 
\newline{
E-Mail: yasu.stat@gmail.com}}, \
Kaoru Irie\footnote{Faculty of Economics, The University of Tokyo. 
\newline{
E-Mail: irie@e.u-tokyo.ac.jp}}, \
and 
Shonosuke Sugasawa\footnote{Center for Spatial Information Science, The University of Tokyo. 
\newline{
E-Mail: sugasawa@csis.u-tokyo.ac.jp}} 
}
\maketitle
\begin{abstract}
In this paper, we introduce a new and efficient data augmentation approach to the posterior inference of the models with shape parameters when the reciprocal gamma function appears in full conditional densities. 
Our approach is to approximate full conditional densities of shape parameters by using Gauss's multiplication formula and Stirling's formula for the gamma function, where the approximation error can be made arbitrarily small. 
We use the techniques to construct efficient Gibbs and Metropolis-Hastings algorithms for a variety of models that involve the gamma distribution, Student's $t$-distribution, the Dirichlet distribution, the negative binomial distribution, and the Wishart distribution. 
The proposed sampling method is numerically demonstrated through simulation studies.

\par\vspace{4mm}
\end{abstract}

\section{Introduction}
\label{sec:introduction}
Markov chain Monte Carlo (MCMC) algorithms are now widely adopted in Bayesian posterior computation, where parameters are iteratively sampled from their respective conditional distributions. 
However, when the models of interest involve the gamma and related distributions, it is computationally costly to sample the shape parameters from their full conditional posteriors. 
The main difficulty here is that the full conditional densities of the shape parameters involve the reciprocal gamma function, $1 / \Ga ( \xi )$, $\xi > 0$, and are not any well-known distributions. 
Thus, it is not straightforward to construct efficient MCMC algorithms when the shape parameters are also estimated. 

Several sampling strategies that have been proposed in the literature are customized for each class of distributions.
For gamma distributions, Miller (2019) provided an accurate approximation of the full conditional distribution of the shape parameter. 
For Student's $t$-distributions, Fonseca et al. (2008) considered the unknown degrees of freedom, at the cost of the complication of the priors. 
For Dirichlet-multinomial and negative-binomial models, sampling algorithms for the shape parameters have been proposed by Nandram (1998) and Zhou and Carin (2015), respectively.

Rather than focusing on a particular class of distributions, it is also possible to devise the sampling methods that are applicable to the general class of models with shape parameters, at the cost of efficiency and computational time. 
For example, the approximation of log-concave densities (Gilks and Wild, 1992; Devroye, 2012) and the MH acceptance-rejection method (Tierney, 1994; Chib and Greenberg, 1995) can be used for the posterior inference for models with the reciprocal gamma functions. The latter needs to be further customized to each model, as practiced for Student's $t$-models in Watanabe (2001). 
Another approach is the data augmentation scheme, where several latent variables are introduced to simplify the full conditionals of the model. 
He et al. (2021) proposed a general and efficient data augmentation for models with reciprocal gamma functions, where the simulation from power truncated normal (PTN) distributions become necessary. In this paper, we also take the data-augmentation approach, but propose a new augmentation where we only need to simulate from well-known distributions.

Our strategy for deriving an augmented model is twofold: (i) using Gauss's multiplication formula for the gamma function to introduce conditionally beta-distributed latent variables and (ii) approximating the augmented densities by Stirling's formula. 
The full conditionals of the shape parameters and latent variables of the resulting model are all well-known distributions, such as gamma and beta distributions, from which it is easy and fast to simulate. Finally, the accept/reject step is added to justify the sampling algorithm as an independent Metropolis-Hastings (MH) method. 

To assess the efficiency of the sampling algorithm based on the proposed augmentation, we evaluate the upper and lower bounds of the approximation error and show that, in many cases, the acceptance probability is close to one. 
Due to its simplicity, our augmentation scheme can be applied directly to many models with reciprocal gamma functions, including the Student's $t$-distribution, Dirichlet-multinomial distribution, negative binomial distribution and Wishart distribution.

The remainder of the paper is organized as follows. 
In Section \ref{sec:general}, we develop a new data augmentation and approximation of the reciprocal gamma function and illustrate our approach using a simple gamma model. 
For simplicity, we consider only proper priors for shape parameters as well as other variables, which ensures that full conditional distributions are always proper. 
In Section \ref{sec:t}, we use our approach for a model based on Student's $t$-distribution. 
In Section \ref{sec:dir2}, we consider a Dirichlet-multinomial model and apply a generic method. 
Some concluding remarks are given in Section \ref{sec:conclusion}. 
Proofs and additional results are provided in the Supplementary Material.

\section{Beta Data Augmentation}
\label{sec:general}

\subsection{General ideas}
\label{subsec:properties_gamma} 
The most important result for our method is the following integral expression, which is based on Gauss’s multiplication formula for the gamma function.

\begin{thm}
\label{cor:2} 
Let $m \in \mathbb{N}$. 
Then we have 
\begin{align}
{1 \over \{ \Ga ( \xi ) \} ^m} &= C_m {1 \over \xi ^{m \xi }} \xi ^{m + 1 / 2 - 1} e^{m \xi } \Big\{ \prod_{j = 2}^{m} \int_{0}^{1} {\rho _j}^{\xi + (j - 1) / m - 1} (1 - \rho _j )^{(m - j + 1) / m - 1} d{\rho _j} \Big\} {(m \xi )^{m \xi - 1 / 2} \over \Ga (m \xi ) e^{m \xi }} ,\non 
\end{align}
for all $\xi > 0$, where $C_m = 1 / \big\{ (2 \pi )^{(m - 1) / 2} \prod_{j = 2}^{m} \Ga ((m - j + 1) / m) \big\} $. 
\end{thm}

\bigskip
The proof is given in the Supplementary Material. By Theorem~\ref{cor:2}, we can rewrite the $m$th power of the reciprocal gamma function, $1 / \{ \Ga ( \xi ) \} ^m$, by using integrals of $m - 1$ beta densities, such that the reciprocal gamma function appears only once in the right-hand side.

Suppose that the target distribution, or the posterior distribution, is the joint density of shape parameter $\xi $ and other variables $\bvth $ of the form,
\begin{align}
p( \xi , \bvth ) &\propto f( \xi , \bvth ) {1 \over \{ \Ga ( \xi ) \} ^m} \text{,} \non 
\end{align}
where typically $f( \xi , \bvth ) \stackrel{\xi }{\propto } {\rm{Ga}} ( \xi | a_1 , b_1 )$ for some $a_1 , b_1 > 0$. 
This framework covers, for example, the case of $n$ independent observations from a gamma distribution, $x_1 , \dots , x_n \sim {\rm{Ga}} ( \al , \be )$; in this case, $m = n$, $( \xi , \bvth ) = ( \al , \be )$, and $p( \al , \be )$ is the posterior of $( \al ,\be )$ given $x_1,\dots , x_n$, or $p( \al , \be ) \propto \pi ( \al ,\be ) \times \be ^{n \al } \big( \prod_{i = 1}^{n} x_i \big) ^{\al } e^{- \be \sum_{i = 1}^{n} x_i}$, where $\pi ( \al ,\be )$ is a prior density (see Section \ref{subsec:ga}). 
In general, some of the variables $\bvth $ may be latent variables introduced based on data augmentation. 
We are interested in the repeated sampling from the conditional distributions, $p(\xi|\bvth)$ and $p(\bvth|\xi)$. We assume that it is relatively easy to sample $\bvth $ from $p(\bvth|\xi)$, and we focus on the problem of sampling $\xi $ from $p(\xi|\bvth)$ in the following. 

The derivation of the augmented model is a three-step process. First, we rewrite $p( \xi , \bvth )$ as 
\begin{align}
p( \xi , \bvth ) &\propto {f( \xi , \bvth ) \over \xi ^{m \xi }} \xi ^{m + 1 / 2 - 1} e^{m \xi } \Big\{ \prod_{j = 2}^{m} \int_{0}^{1} {\rho _j}^{\xi + (j - 1) / m - 1} (1 - \rho _j )^{(m - j + 1) / m - 1} d{\rho _j} \Big\} {(m \xi )^{m \xi - 1 / 2} \over \Ga (m \xi ) e^{m \xi }} ,\non 
\end{align}
by using Theorem \ref{cor:2}. 
The $m$th power, $1 / \{ \Ga ( \xi ) \} ^m$, is simplified to a single reciprocal gamma function, $1 / \Ga ( m\xi )$, which we further evaluate in the following steps. 
A set of additional latent variables, $\bro = ( \rho _2 , \dots , \rho _m ) \in (0, 1)^{m - 1}$, has the full conditional of the simple form, $\prod_{j = 2}^{m} {\rm{Beta}} ( \rho _j | \xi + (j - 1) / m, (m - j + 1) / m)$, from which we can easily sample.

Second, the conditional density of $( \xi , \bvth )$ given $\bro $ is 
\begin{align}
p( \xi , \bvth | \bro ) &\propto {f( \xi , \bvth ) \over \xi ^{m \xi }} \xi ^{a_2 - 1} e^{- b_2 \xi } {(m \xi )^{m \xi - 1 / 2} \over \Ga (m \xi ) e^{m \xi }} \text{,} \non 
\end{align}
where $a_2 = m + 1 / 2$ and $b_2 = - m + \sum_{j = 2}^{m} \log (1 / \rho _j )$. 
In the above expression, there are two factors that make it difficult to sample $\xi $ from the full conditional: $1 / \xi ^{m \xi }$ and $(m \xi )^{m \xi - 1 / 2} / \{ \Ga (m \xi ) e^{m \xi } \} $. 
Here, in order to eliminate $1 / \xi ^{m \xi }$, we assume that we can make the change of variables $\widetilde{\bvth } = \varphi ( \bvth ; \xi )$ with Jacobian $\xi ^{m \xi }$, so that $f( \xi , \bvth ) d(\xi , \bvth) = \xi ^{m \xi } \ft ( \xi , \widetilde{\bvth } ) d(\xi , \widetilde{\bvth })$, and the density of interest becomes 
\begin{align}
p( \xi , \widetilde{\bvth } | \bro ) &\propto \ft ( \xi , \widetilde{\bvth } ) \xi ^{a_2 - 1} e^{- b_2 \xi } {(m \xi )^{m \xi - 1 / 2} \over \Ga (m \xi ) e^{m \xi }} \text{.} \non 
\end{align}
This change-of-variable is available for many models, including the gamma model of Section \ref{subsec:ga}. The models for which there is no such change-of-variable, including the Dirichlet-multinomial model of Section \ref{sec:dir2}, are discussed in Section~\ref{subsec:generic_method}.

Third, we use the above expression to construct an independent MH algorithm. 
Let $\xi ^{\rm{old}}$ be a current value of $\xi $. 
To generate a new value $\xi ^{\rm{new}}$, we first sample a proposal $\xi ^{*}$ from the approximate full conditional density proportional to $\ft ( \xi , \widetilde{\bvth } ) \xi ^{a_2 - 1} e^{- b_2 \xi }$ and compute 
\begin{align}
p = \min \Big\{ 1, {(m \xi ^{*} )^{m \xi ^{*} - 1 / 2} \over \Ga (m \xi ^{*} ) e^{m \xi ^{*} }} / {(m \xi ^{\rm{old}} )^{m \xi ^{\rm{old}} - 1 / 2} \over \Ga (m \xi ^{\rm{old}} ) e^{m \xi ^{\rm{old}}}} \Big\} \text{.} \non 
\end{align}
Then we set $\xi ^{\rm{new}} = \xi ^{*}$ with probability $p$, otherwise $\xi ^{\rm{new}} = \xi ^{\rm{old}}$. 
We note that in all the models considered in this paper, proposal distributions corresponding to $\ft ( \xi , \widetilde{\bvth } ) \xi ^{a_2 - 1} e^{- b_2 \xi }$ are easy to sample from. 
The factor dropped in the approximate distribution can be evaluated as 
\begin{align}
&{e^{- 1 / (12 \xi )} \over (2 \pi )^{1 / 2}} < {\xi ^{\xi - 1 / 2} \over \Ga ( \xi ) e^{\xi }} < {1 \over (2 \pi )^{1 / 2}} \label{eq:stir} ,
\end{align}
for any $\xi > 0$ by Stirling's formula. This expression shows that the factor is almost constant when $\xi$ is not extremely small, and that the acceptance probability $p$ is close to one. 
This can be confirmed by bounding the acceptance probability below as $p \ge e^{- 1 / (12 m \xi ^{*} )} \ge 1 - 1 / (12 m \xi ^{*} )$, where the lower bound is almost unity unless $\xi^{*}$ is extremely small.

\subsection{An illustration using a gamma model}
\label{subsec:ga} 
Here, we consider a simple gamma model for illustration. 
For this model, several methods for posterior inference are available (e.g.~Gilks and Wild 1992).  In particular, the method of Miller (2019) is customized for this model and highly efficient. 

Suppose that observations $\x = ( x_1 , \dots , x_n )$ have been independently generated from a gamma distribution ${\rm{Ga}} ( \al , \be )$. 
We assume the independent gamma prior distributions for $\alpha$ and $\beta$: ${\rm{Ga}} (a, b)$ and ${\rm{Ga}} (c, d)$, respectively.
Then the posterior of $( \al , \be )$ is 
\begin{align}
p( \al , \be | \x ) &\propto {\rm{Ga}} ( \al | a, b) \be ^{c - 1} e^{- d \be } {\be ^{n \al } \over \{ \Ga ( \al ) \} ^n} \Big( \prod_{i = 1}^{n} x_i \Big) ^{\al } e^{- \be \sum_{i = 1}^{n} x_i} \text{.} \non 
\end{align}
Using Theorem \ref{cor:2}, we can rewrite the above posterior density as 
\begin{align}
p( \al , \be | \x ) &\propto {\rm{Ga}} ( \al | a, b) \be ^{c - 1} e^{- d \be } \be ^{n \al } \Big( \prod_{i = 1}^{n} x_i \Big) ^{\al } e^{- \be \sum_{i = 1}^{n} x_i} \non \\
&\quad \times {1 \over \al ^{n \al }} \al ^{n + 1 / 2 - 1} e^{n \al } \Big\{ \prod_{i = 2}^{n} \int_{0}^{1} {\rho _i}^{\al + (i - 1) / n - 1} (1 - \rho _i )^{(n - i + 1) / n - 1} d{\rho _i} \Big\} {(n \al )^{n \al - 1 / 2} \over \Ga (n \al ) e^{n \al }} \text{.} \non 
\end{align}
Now, we consider $\bro = ( \rho _2 , \dots , \rho _n ) \in (0, 1)^{n - 1}$ as a set of additional latent variables. 
Then the conditional distribution of $( \al , \be , \bro )$ given $\x $ is 
\begin{align}
p( \al , \be , \bro | \x ) &\propto {\rm{Ga}} ( \al | a, b) \be ^{c - 1} e^{- d \be } \be ^{n \al } \Big( \prod_{i = 1}^{n} x_i \Big) ^{\al } e^{- \be \sum_{i = 1}^{n} x_i} \non \\
&\quad \times {1 \over \al ^{n \al }} \al ^{n + 1 / 2 - 1} e^{n \al } \Big[ \prod_{i = 2}^{n} \{ {\rho _i}^{\al + (i - 1) / n - 1} (1 - \rho _i )^{(n - i + 1) / n - 1} \} \Big] {(n \al )^{n \al - 1 / 2} \over \Ga (n \al ) e^{n \al }} \text{.} \non 
\end{align}

In order to obtain MCMC samples of $( \al , \be , \bro ) | \x $, we can use the MH within Gibbs sampler. 
It is easy to sample $\bro $ from its full conditional distribution since $p( \bro | \al , \be , \x ) = \prod_{i = 2}^{n} {\rm{Beta}} ( \rho _i | \al + (i - 1) / n, (n - i + 1) / n)$. 
Meanwhile, the full conditional of $( \al , \be )$ is 
\begin{align}
p( \al , \be | \bro , \x ) &\propto {1 \over \al ^{n \al }} \al ^{n - 1 / 2 + a - 1} \exp \Big\{ - \al \Big( - \sum_{i = 1}^{n} \log x_i + \sum_{i = 2}^{n} \log {1 \over \rho _i} - n + b \Big) \Big\} \non \\
&\quad \times \be ^{n \al + c - 1} \exp \Big\{ - \be \Big( \sum_{i = 1}^{n} x_i + d \Big) \Big\} {(n \al )^{n \al - 1 / 2} \over \Ga (n \al ) e^{n \al }} \text{.} \non 
\end{align}
Although the full conditional of $\be $ is a gamma distribution, the full conditional density of $\al $ does not have a standard form because of the two factors: $g_1 ( \al ) = 1 / \al ^{n \al }$ and $g_2 ( \al ) = (n \al )^{n \al - 1 / 2} / \{ \Ga (n \al ) e^{n \al } \} $. 

First, in order to eliminate $g_1 ( \al )$ from the above expression, we make the change of variables $\ga = \be / \al $. 
Then 
\begin{align}
p( \al , \ga | \bro , \x ) &\propto \al ^{n - 1 / 2 + c + a - 1} \exp \Big\{ - \al \Big( - \sum_{i = 1}^{n} \log x_i + \sum_{i = 2}^{n} \log {1 \over \rho _i} - n + b \Big) \Big\} \non \\
&\quad \times \ga ^{n \al + c - 1} \exp \Big\{ - \al \ga \Big( \sum_{i = 1}^{n} x_i + d \Big) \Big\} g_2 ( \al ) \text{.} \non 
\end{align}
The full conditional of $\ga = \be / \al $ is given by ${\rm{Ga}} \big( \ga \big| n \al + c, \al \big( \sum_{i = 1}^{n} x_i + d \big) \big) $ and tractable similar to that of the original parameter $\be $. 

Next, we use the MH algorithm to update $\al $. 
The full conditional density of $\al$ is given by $p( \al | \ga , \bro , \x ) \propto {\rm{Ga}} ( \al | A, B) g_2 ( \al )$, where $A = n - 1 / 2 + c + a$ and $B = - \sum_{i = 1}^{n} \log x_i + \sum_{i = 2}^{n} \log (1 / \rho _i ) - n - n \log \ga + \ga \big( \sum_{i = 1}^{n} x_i + d \big) + b$. 
We sample a proposal $\al ^{*}$ from ${\rm{Ga}} ( \al | A, B)$. 
We accept $\al ^{*}$ if an independent standard uniform variable $U$ is less than or equal to $g_2 ( \al ^{*} ) / g_2 ( \al ^{\rm{old}} )$, where $\al ^{\rm{old}}$ denotes the current value of $\al $. 
The new value of $\al $, or $\al ^{\rm{new}}$, is set to $\al ^{*}$ if $\al ^{*}$ is accepted, and to $\al ^{\rm{old}}$ otherwise. 

The MH within Gibbs sampler is summarized as follows. 
\begin{algo}
The variables $\al $, $\ga $, and $\bro $ are updated in the following way. 
\begin{itemize}
\item[-]
Sample $\ga ^{*} \sim {\rm{Ga}} \big( n \al + c, \al \big( \sum_{i = 1}^{n} x_i + d \big) \big) $. 
\item[-]
Sample $\bro ^{*} = ( \rho _{2}^{*} , \dots , \rho _{n}^{*} ) \sim \prod_{i = 2}^{n} {\rm{Beta}} ( \al + (i - 1) / n, (n - i + 1) / n)$. 
\item[-]
Sample $\al ^{*} \sim {\rm{Ga}} (A, B)$, where $A = n - 1 / 2 + c + a$ and 
\begin{align}
B = - \sum_{i = 1}^{n} \log x_i + \sum_{i = 2}^{n} \log {1 \over \rho _{i}^{*}} - n - n \log \ga ^{*} + \ga ^{*} \Big( \sum_{i = 1}^{n} x_i + d \Big) + b \text{,} \non 
\end{align}
and accept $\al ^{*}$ with probability 
\begin{align}
\min \Big\{ 1, {(n \al ^{*} )^{n \al ^{*} - 1 / 2} \over \Ga (n \al ^{*} ) e^{n \al ^{*}}} / {(n \al )^{n \al - 1 / 2} \over \Ga (n \al ) e^{n \al }} \Big\} \text{.} \non 
\end{align}
\end{itemize}
\end{algo}

The accuracy of approximation, or the acceptance probability, has already been evaluated in (\ref{eq:stir}). The acceptance probability is, at least, $1 - 1 / (12 n \al ^{*} )$.

\subsection{PTN data augmentation}
\label{subsec:generic_method}

The key to the augmentation strategy of Section \ref{subsec:properties_gamma} is to find suitable changes of variables $\widetilde{\bvth } = \varphi ( \bvth ; \xi )$ to eliminate the factor $1 / \xi ^{m \xi }$ in the second step. 
Because this is not always straightforward, an alternative method is developed in this section. 
We modify the proposed method of Section \ref{subsec:properties_gamma} by introducing additional latent variables. 
The main tool is the integral expression in the following lemma.

\begin{lem}
\label{lem:generic} 
Let $m \in \mathbb{N}$. 
Then 
\begin{align}
{1 \over {\xi }^{m \xi }} &= (m \xi )^{1 / 2} e^{m \xi } {(m \xi )^{m \xi - 1 / 2} \over \Ga (m \xi ) e^{m \xi }} \int_{0}^{\infty } w^{m \xi - 1} e^{- w m {\xi }^2} dw \non 
\end{align}
for all $\xi > 0$. 
\end{lem}

We assume that $f( \xi , \bvth ) \stackrel{\xi }{\propto } {\rm{Ga}} ( \xi | a_1 , b_1 )$ for simplicity and consider the conditional density 
\begin{align}
p( \xi | \bro , \bvth ) &\propto {1 \over \xi ^{m \xi }} \xi ^{a_3 - 1} e^{- b_3 \xi } {(m \xi )^{m \xi - 1 / 2} \over \Ga (m \xi ) e^{m \xi }} \text{,} \non 
\end{align}
where $a_3 = a_1 + m - 1 / 2$ and $b_3 = b_1 - m + \sum_{j = 2}^{m} \log (1 / \rho _j )$. 
Using Lemma \ref{lem:generic}, we see that $p( \xi | \bro , \bvth )$ is the marginal density of 
\begin{align}
p( \xi , w | \bro , \bvth ) &\propto \xi ^{a_3 - 1 / 2} e^{- ( b_3 - m) \xi } \Big\{ {(m \xi )^{m \xi - 1 / 2} \over \Ga (m \xi ) e^{m \xi }} \Big\} ^2 w^{m \xi - 1} e^{- w m {\xi }^2} \text{,} \non 
\end{align}
where $w \in (0, \infty )$ is an additional latent variable. 
Clearly, $p( w | \xi , \bro , \bvth ) = {\rm{Ga}} (w | m \xi , m \xi ^2 )$. 
On the other hand, 
\begin{align}
p( \xi | w, \bro , \bvth ) / \Big\{ {(m \xi )^{m \xi - 1 / 2} \over \Ga (m \xi ) e^{m \xi }} \Big\} ^2 &\propto \xi ^{c - 1} e^{- a {\xi }^2 + b \xi } \text{,} \label{eq:full_conditional_PTN} 
\end{align}
where $c = a_3 + 1 / 2$, $a = m w$, and $b = m \log w + m - b_3$. 
The right-hand side is proportional to the power truncated normal (PTN) distribution (He et al., 2021) with parameters $c$, $a$, and $b$, which is denoted by ${\rm{PTN}} (c, a, b)$. 
Since the denominator of the left-hand side in (\ref{eq:full_conditional_PTN}) is almost constant as seen in (\ref{eq:stir}), the conditional density $p( \xi | w, \bro , \bvth )$ is approximated by ${\rm{PTN}} ( \xi | c, a, b)$. 
Then, we generate a proposal, $\xi ^{*} \sim {\rm{PTN}} (c, a, b)$, and accept it with probability 
\begin{align}
\min \Big\{ 1, \Big\{ {(m \xi ^{*} )^{m \xi ^{*} - 1 / 2} \over \Ga (m \xi ^{*} ) e^{m \xi ^{*}}} / {(m \xi ^{\rm{old}} )^{m \xi ^{\rm{old}} - 1 / 2} \over \Ga (m \xi ^{\rm{old}} ) e^{m \xi ^{\rm{old}}}} \Big\} ^2 \Big\} \text{,} \non 
\end{align}
where $\xi ^{\rm{old}}$ is the current state of $\xi $.

\subsection{Additional data augmentation for the PTN distribution}
\label{subsec:additional_augmentation} 
In order to sample from the PTN distribution (\ref{eq:full_conditional_PTN}), one can use the accept/reject algorithm described by He et al. (2021). 
In this paper, we consider other approaches so that we do not necessarily need to use accept/reject algorithms. 
Our approaches also have potential flexibility that they are easily extended to the case where $f( \xi , \bvth )$ is proportional to a generalized-inverse-Gaussian density as a function of $\xi $. 

Let $M > 0$ be a constant possibly dependent on $w$, $\bro $, and $\bvth $ such that $M > b$. 
(A convenient choice is $M = 1 + \max \{ 0, b \} $.) 
Then, the PTN density is written as 
\begin{align}
{\rm{PTN}} ( \xi | c, a, b) &\propto \xi ^{c - 1} e^{- a {\xi }^2 - b' \xi } e^{M \xi } \text{,} \non 
\end{align}
where $c$, $a$, $b' = M - b$, and $M$ are all positive. 

The exponential term $e^{M \xi }$ can be augmented in two ways. The first approach is based on the following expression: 
\begin{align}
{\rm{PTN}} ( \xi | c, a, b) &\propto \xi ^{c - 1} e^{- a {\xi }^2 - b' \xi } \sum_{\ze = 0}^{\infty } {M^{\ze } \xi ^{\ze } \over \ze !} \non \\
&= \sum_{\ze = 0}^{\infty } {M^{\ze } \xi ^{\ze } \over \ze !} \xi ^{c - 1} e^{- a {\xi }^2} \int_{0}^{\infty } {1 \over \sqrt{2 \pi }} \eta ^{1 / 2 - 1} e^{- \eta / 2} e^{- ( b' )^2 \xi ^2 / (2 \eta )} d\eta \text{,} \non 
\end{align}
where we consider $\ze \in \mathbb{N} _0 = \{ 0, 1, 2, \dotsc \} $ and $\eta \in (0, \infty )$ as additional latent variables. 
Then the full conditional distributions of $\ze $ and $\eta $ are ${\rm{Po}} ( \ze | M \xi )$ and ${\rm{GIG}} ( \eta | 1 / 2, 1, ( b' )^2 \xi ^2 )$, respectively. 
The full conditional density of $\xi $ divided by $\{ (m \xi )^{m \xi - 1 / 2} / \Ga (m \xi ) e^{m \xi } \} ^2$ is proportional to 
\begin{align}
\xi ^{\ze + c - 1} e^{- \{ a + ( b' )^2 / (2 \eta ) \} {\xi }^2 } \text{.} \non 
\end{align}
We can easily sample from the above distribution since it is simply the square root of a gamma variable. 

The second approach utilizes the integral expression based on the normal density. 
\begin{lem}
\label{lem:normal} 
For all $\xi > 0$, we have 
\begin{align}
e^{\xi } &= \int_{- \infty }^{\infty } {1 \over \sqrt{2 \pi }} {1 \over \sqrt{2 \xi }} e^{- \th ^2 / (4 \xi ) +\th } d\th \text{.} \non 
\end{align}
\end{lem}
By this lemma, we have
\begin{align}
{\rm{PTN}} ( \xi | c, a, b) &\propto \xi ^{c - 1} e^{- a {\xi }^2 - b' \xi } \int_{- \infty }^{\infty } {\xi }^{1 / 2 - 1} e^{- \th ^2 / (4 M \xi ) + \th } d\th \non \\
&\propto \int_{- \infty }^{\infty } e^{\th } \xi ^{c - 1 / 2 - 1} e^{- a {\xi }^2} \Big[ \int_{0}^{\infty } \eta ^{1 / 2 - 1} e^{- \eta / 2} e^{- \{ b' \xi + \th ^2 / (4 M \xi ) \} ^2 / (2 \eta )} d\eta \Big] d\th \text{,} \non 
\end{align}
where we consider $\th \in (- \infty , \infty )$ and $\eta \in (0, \infty )$ as additional latent variables. 
Sampling from the full conditional of $(\eta , \th)$ can be done in a compositional way; we sample $\th$ (with $\eta$ marginalized out) from ${\rm{N}} ( \th | 2 M \xi , 2 M \xi )$, then $(\eta |\th)$ from ${\rm{GIG}} ( \eta | 1 / 2, 1, \{ b' \xi + \th ^2 / (4 M \xi ) \} ^2 )$. 
The full conditional density of $\xi $ divided by $\{ (m \xi )^{m \xi - 1 / 2} / \Ga (m \xi ) e^{m \xi } \} ^2$ is proportional to 
\begin{align}
\xi ^{c - 1 / 2 - 1} e^{- \{ a + ( b' )^2 / (2 \eta ) \} \xi ^2} e^{- \{ \th ^4 / (32 M^2 \eta ) \} / \xi ^2} \text{,} \non 
\end{align}
which is the square root of a generalized-inverse-Gaussian distribution.

\section{Student's $t$-Distribution}
\label{sec:t} 
\subsection{Sampling algorithm}
\label{subsec:t_algorithm}

Student's $t$-distribution is widely adopted in Bayesian inference to handle outliers in samples or heavy-tailed properties of data generating processes (e.g. Geweke, 1993; Fonseca et al., 2008; Villa and Rubio, 2018; da Silva et al., 2020).
A typical problem in using Student's $t$-distribution is that the posterior inference of the degrees of freedom is not straightforward since its full conditional distribution has a complicated form. 
However, we can use our data-augmentation approach. 
We here consider the simplest case where the means of all observations are the same, and we use the normal-scale-mixture representation of Student's $t$-distribution, under which the degrees-of-freedom parameter is regarded as the shape parameter in the gamma distribution.

Suppose that for $i = 1, \dots , n$, 
\begin{align}
&x_i \sim {\rm{t}} ( x_i | ( \th , \ta ), 2 \al ) = {\Ga ( \al + 1 / 2) \over \sqrt{2 \pi } \ta ^{1 / 2} \al ^{1 / 2} \Ga ( \al )} / \Big\{ 1 + {( x_i - \th )^2 \over 2 \ta \al } \Big\} ^{\al + 1 / 2} \text{,} \non \\
&( \th , \ta ) \sim p( \th , \ta ) \text{,} \ \ \ \ \ 
\al \sim p( \al ) \text{,} \non 
\end{align}
where $x_i \in \mathbb{R}$, $\th \in \mathbb{R}$, and $\ta \in (0, \infty )$. 
Then the posterior distribution $p( \th , \ta , \al | \x )$ is obtained as the marginal distribution of 
\begin{align}
p( \th , \ta , \al , \w , \bro | \x ) &\propto {p( \th , \ta ) \over \ta ^{n / 2}} p( \al ) \al ^{n + 1 / 2 - 1} e^{n \al } \Big( \prod_{i = 1}^{n} [ {w_i}^{\al + 1 / 2 - 1} e^{- w_i \{ \al + ( x_i - \th )^2 / (2 \ta ) \} } ] \Big) \non \\
&\quad \times \Big[ \prod_{i = 2}^{n} \{ {\rho _i}^{\al + (i - 1) / n - 1} (1 - \rho _i )^{(n - i + 1) / n - 1} \} \Big] {(n \al )^{n \al - 1 / 2} \over \Ga (n \al ) e^{n \al }} \text{,} \label{eq:t_prp} 
\end{align}
where $\w = ( w_1 , \dots , w_n ) \in (0, \infty )^n$ and $\bro = ( \rho _2 , \dots , \rho _n ) \in (0, 1)^{n - 1}$ are additional latent variables. 
The above expression is derived in Section 
S6 of the Supplementary Material by using Theorem \ref{cor:2}. 

If we use the priors $p( \th , \ta ) = {\rm{N}} ( \th | b, \ta / a) {\rm{IG}} ( \ta | c, d)$ and $p( \al ) = {\rm{Ga}} ( \al | a_0 , b_0 )$ for $a, c, d \in (0, \infty )$ and $b \in \mathbb{R}$ and $a_0 , b_0 \in (0, \infty )$, we can use the following algorithm to generate posterior samples. 
\begin{algo}
\label{algo:t} 
The variables $\th $, $\ta $, $\al $, $\w $, and $\bro $ are updated in the following way. 
\begin{itemize}
\item[-]
Sample $\ta ^{*} \sim {\rm{IG}} ( c' , d' )$, where $c' = n / 2 + c$ and 
\begin{align}
d' &= {1 \over 2} \Big\{ a b^2 + \sum_{i = 1}^{n} w_i {x_i}^2 - {\big( a b + \sum_{i = 1}^{n} w_i x_i \big) ^2 \over a + \sum_{i = 1}^{n} w_i} \Big\} + d \text{.} \non 
\end{align}
\item[-]
Sample $\th ^{*} \sim {\rm{N}} ( b' , \ta ^{*} / a' )$, where $a' = a + \sum_{i = 1}^{n} w_i$ and 
\begin{align}
b' &= {a b + \sum_{i = 1}^{n} w_i x_i \over a + \sum_{i = 1}^{n} w_i} \text{.} \non 
\end{align}
\item[-]
Sample $\w ^{*} = ( w_{1}^{*} , \dots , w_{n}^{*} ) \sim \prod_{i = 1}^{n} {\rm{Ga}} ( \al + 1 / 2, \al + ( x_i - \th ^{*} )^2 / (2 \ta ^{*} ))$. 
\item[-]
Sample $\bro ^{*} = ( \rho _{2}^{*} , \dots , \rho _{n}^{*} ) \sim \prod_{i = 2}^{n} {\rm{Beta}} ( \al + (i - 1) / n, (n - i + 1) / n)$. 
\item[-]
Sample $\al ^{*} \sim {\rm{Ga}} ( {a_0}' , {b_0}' )$, where ${a_0}' = a_0 + n - 1 / 2$ and 
\begin{align}
{b_0}' &= b_0 - n + \sum_{i = 1}^{n} ( w_{i}^{*} - \log w_{i}^{*} ) + \sum_{i = 2}^{n} \log {1 \over \rho _{i}^{*}} \text{,} \non 
\end{align}
and accept $\al ^{*}$ with probability 
\begin{align}
\min \Big\{ 1, {(n \al ^{*} )^{n \al ^{*} - 1 / 2} \over \Ga (n \al ^{*} ) e^{n \al ^{*}}} / {(n \al )^{n \al - 1 / 2} \over \Ga (n \al ) e^{n \al }} \Big\} \text{.} \non 
\end{align}
\end{itemize}
\end{algo}

Since we introduce the additional latent variables $\rho _2 , \dots , \rho _n$, our method is less efficient than an alternative method in terms of the effective sample size for an MCMC sequence of a fixed number of parameter values. 
However, since we do not need to use numerical approximation, our method takes less time. 
These are confirmed in Section \ref{subsec:t_sim}. 

We remark that our method is flexible and we can use many other types of priors. 
For example, we can use a scale mixture of gamma distributions as a prior for $\al $. 
We can use a truncated gamma prior for $\al $ and this case is considered in the second half of Section \ref{subsec:t_sim}. 
Also, for $a_0 , b_0 , c_0 \in (0, \infty )$, we can use the beta-type prior $p( \al ) \propto \al ^{a_0 - 1} (1 - \al / c_0 )^{b_0 - 1} \chi _{(0, c_0 )} ( \al )$.

\subsection{Simulation study} 
\label{subsec:t_sim} 
Here, we compare the performance of our method based on data augmentation (DA) with the performance of an alternative method based on the approximation proposed by Miller (2019) (A-MH). 
See Section 
S7 of the Supplementary Material for details of the A-MH method. 

First, we set either $n = 10$, $n = 30$, or $n = 100$ and use the conjugate prior $p( \th , \ta ) = {\rm{N}} ( \th | 0, \ta / (1 / 10)) \times {\rm{IG}} ( \ta | 1 / 10, 1 / 10)$ and the gamma prior $p( \al ) = {\rm{Ga}} ( \al | 1 / 10, 1 / 10)$. 
We generate $x_i$ from ${\rm{t}} ( x_i | (3, 1), 2 \al _0 )$. 
We consider the cases $2 \al _0 = 1 / 10$, $2 \al _0 = 1$, and $2 \al _0 = 10$. 
Then, for each of the two methods, we generate $4,000$ posterior samples after discarding the first $1,000$ samples. 
We use $( \ep , M) = ( {10}^{- 8} , 10)$ for the convergence tolerance and the maximum number of iterations for the A-MH method as recommended in Miller (2019). 
We repeat this simulation $100$ times.

Boxplots of the ratios of the effective sample sizes for $\al $, $\ta $, and $\th $ to the computation times for the two methods are shown in Figure \ref{fig:t_1} for $n=10$. 
(For the boxplots for $n=30$ and $n=100$, see Figures S1 and S2 of the Supplementary Material.) 
Table \ref{table:t} reports the averages over the simulations of the ratios (sESS) and the original effective sample sizes (ESS), as well as the mean squared error (MSE) ratios of the estimators of $\al $, $\ta $, and $\th $, where the MSE ratio is defined as the MSE of the alternative method divided by that of our proposed method. 
In terms of MSE, there is little difference between the two methods in many cases including those in the Supplementary Material. 
In terms of sESS, our method is better especially for $\th $ and $\ta $ when $n = 10$. 
When $n = 30$, the alternative method becomes better in terms of $\al $ and competitive in terms of $\ta $ and $\th $. 
When $n = 100$, the alternative method is clearly better than ours. This increase of sESS of the DA method for large $n$ is most likely due to the increased number of latent parameters $\rho _{2:n}$, affecting both efficiency and computational time. For example, the ESSs of center parameter $\th$ are almost unchanged (or even improve) when $n$ increases from $30$ to $100$, hence the decrease of the sESSs for $\th$ is mainly due to the increased computational time.

\begin{figure}[!htbp]
\centering
\includegraphics[width = 16cm]{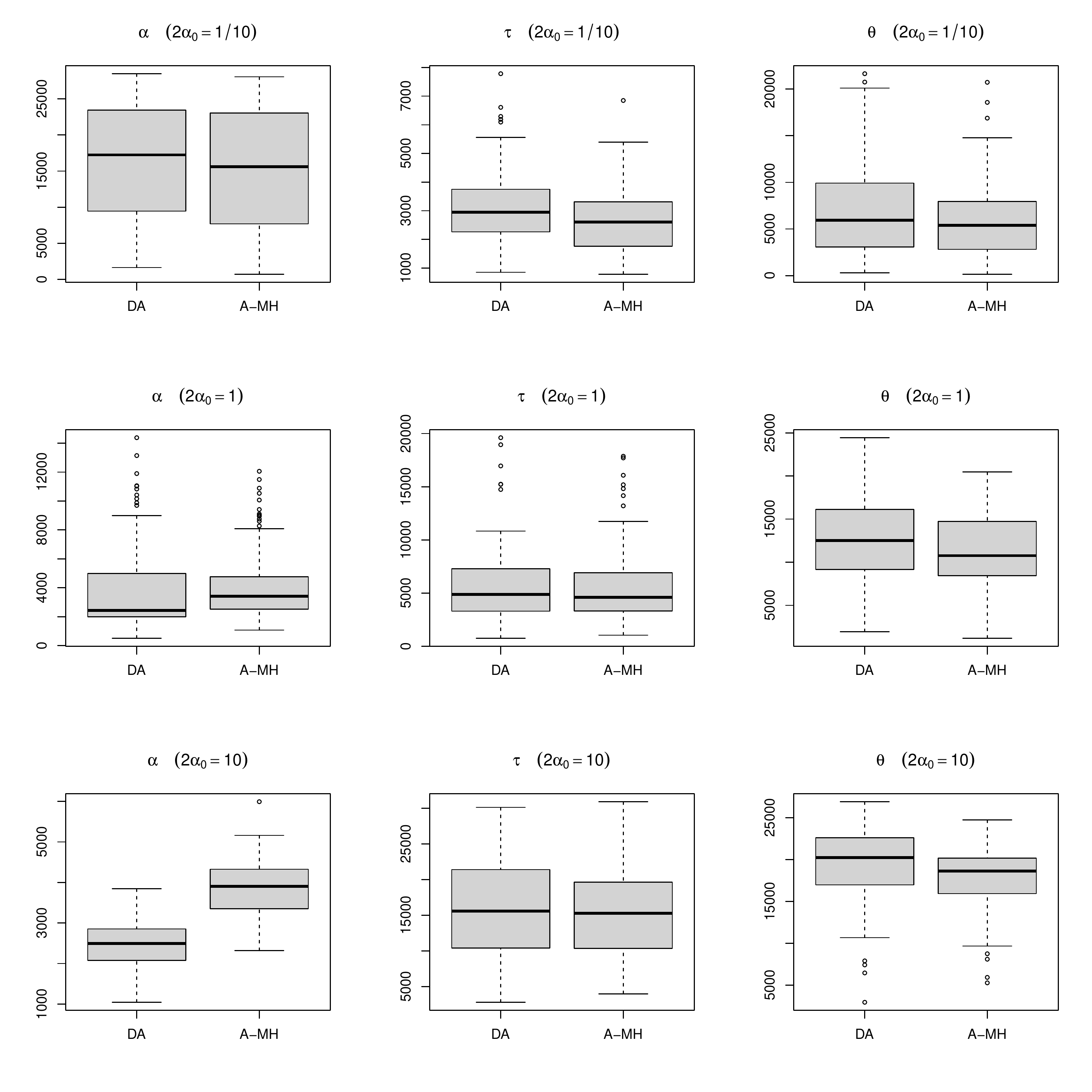}
\caption{Boxplots of the effective sample sizes standardized by the computation times for the proposed method (DA) and the alternative method (A-MH) for $n = 10$. }
\label{fig:t_1}
\end{figure}%

\small
\begin{table}[!thb]
\caption{The averages of the effective sample sizes (ESS) %
for the proposed method (DA) and the alternative method (A-MH) by Miller (2019), the averages of those standardized by computation time (sESS), and the ratios of the mean squared errors (MSE) of the A-MH method to those of the DA method. }
\begin{center}
$
{\renewcommand\arraystretch{1.1}\small
\begin{array}{cccccccccccccccccccc}
\hline &&&& \multicolumn{3}{c}{\text{ESS}} && \multicolumn{3}{c}{\text{sESS}} && \multicolumn{3}{c}{\text{MSE ratio}} \\          
n & 2 \al _0 & \text{method} 
&& \text{$\th $} & \text{$\ta $}
& \text{$\al $} 
&& \text{$\th $} & \text{$\ta $}
& \text{$\al $}
&& \text{$\th $} & \text{$\ta $}
& \text{$\al $}
\\

\hline

10 & 	0.1 & \text{DA} &&	874 & 	382	 & 1972&& 	7148	 & 3121	 & 16058	& & -	 & - & 	-\\

10 & 	0.1 & \text{A-MH} &&		805	 & 361	 & 2097	 && 5887 & 2643 & 	15408 && 	1.14 & 	8.87	 & 1.06 \\

10 & 	1 & \text{DA} &&		1506 & 	686 & 	466& & 	12749& 	5815 & 3926 && 	- & 	-	 & -\\

10 & 	1	 & \text{A-MH} &&	1514 & 	755 	 & 581  && 	11252 & 	5648 & 	4285  && 1.00  & 	1.05 	 & 0.93 \\

10 & 	10 & \text{DA} &&		2276 	 & 1862 & 	288 	 && 19406	 & 15875	 & 2458  && 	-& 	- & - \\

10 & 	10 & \text{A-MH} &&		2316  & 1989 & 	505 && 17634  & 	15193	 & 3854  && 	1.00 	 & 1.01	 & 0.84\\

\hline

30	 & 0.1 & \text{DA} &&		796  & 188 & 	2068 	 && 4836	 & 1135  & 	12525  && 	-& 	- & - \\

30	 & 0.1	 & 	\text{A-MH} && 834  & 	178  & 	2103  && 5115	 & 1095 	 & 12914 && 	1.10  & 	0.79 	 & 1.02  \\

30 & 	1 & \text{DA} &&		1408 & 444 & 	510	 && 9019	 & 2837 	 & 3252 && -	 & -& 	- \\

30	 & 1 & \text{A-MH} &&		1440  & 	473  & 645	 && 9114  & 	2991  & 	4072 	 && 1.00 & 	1.00	 & 1.00 \\

30	 & 10 & \text{DA} &&		2371 	 & 715	 & 122	 && 15235  & 4570 & 	782 	 && -& 	- & -\\

30	 & 10 & \text{A-MH} &&		2448& 921 	 & 220  && 16451 	 & 6235  & 1479 && 1.02	 & 1.02  & 	1.03  \\

\hline

100 & 	0.1 & \text{DA} &&		904 & 	109 & 	1770 && 	3200 	 & 384  & 	6270&& 	- & - & - \\

100	 & 0.1	 & \text{A-MH} &&	895 	 & 105 & 	1802  && 	4040 	 & 472  & 	8120 && 1.00 	 & 0.77	 & 1.00 \\

100	 & 1 & \text{DA} &&		1359 	 & 386 	 & 527	 && 4889& 	1388  & 1888  && -& 	-	 & -\\

100 & 1 & \text{A-MH} &&		1364 & 	391 & 	640 	 && 6251 	 & 1794& 	2926 && 	1.02 & 	1.01 & 1.03 \\

100 & 	10 & \text{DA} &&		2711 & 	284	 & 56 	 && 10380 	 & 1093 	 & 214  && -& 	- & 	- \\

100	 & 10 & \text{A-MH} &&		2721& 	407  & 	100 	 && 13793 	 & 2075  & 	507 && 	1.01 & 	1.04  & 0.98 \\
\hline
\end{array}
}
$
\end{center}
\label{table:t} 
\end{table}
\normalsize

Thus, when $n$ is large, our method benefits rather from its simplicity and applicability to more complicated models. 
To see this point, we consider additional scenarios where a truncated gamma prior is used for the shape parameter; $p( \al ) \propto {\rm{Ga}} ( \al | 1 / 10, 1 / 10) \chi _{( \underline{\al } , \infty )} ( \al )$, where $\underline{\al } > 0$. With this truncated priors, the method of Miller (2019) must evaluate the expected values of truncated gamma distributions, taking longer time for posterior computation. In contrast, no complication is needed for our method to use the truncated prior, except that we now need to sample from truncated distributions. We set $2 \al _0 = 10$ and conduct the same simulation study for the truncated gamma prior with $2 \underline{\al } = 1, 3$. 

\begin{figure}[!htbp]
\centering
\includegraphics[width = 16cm]{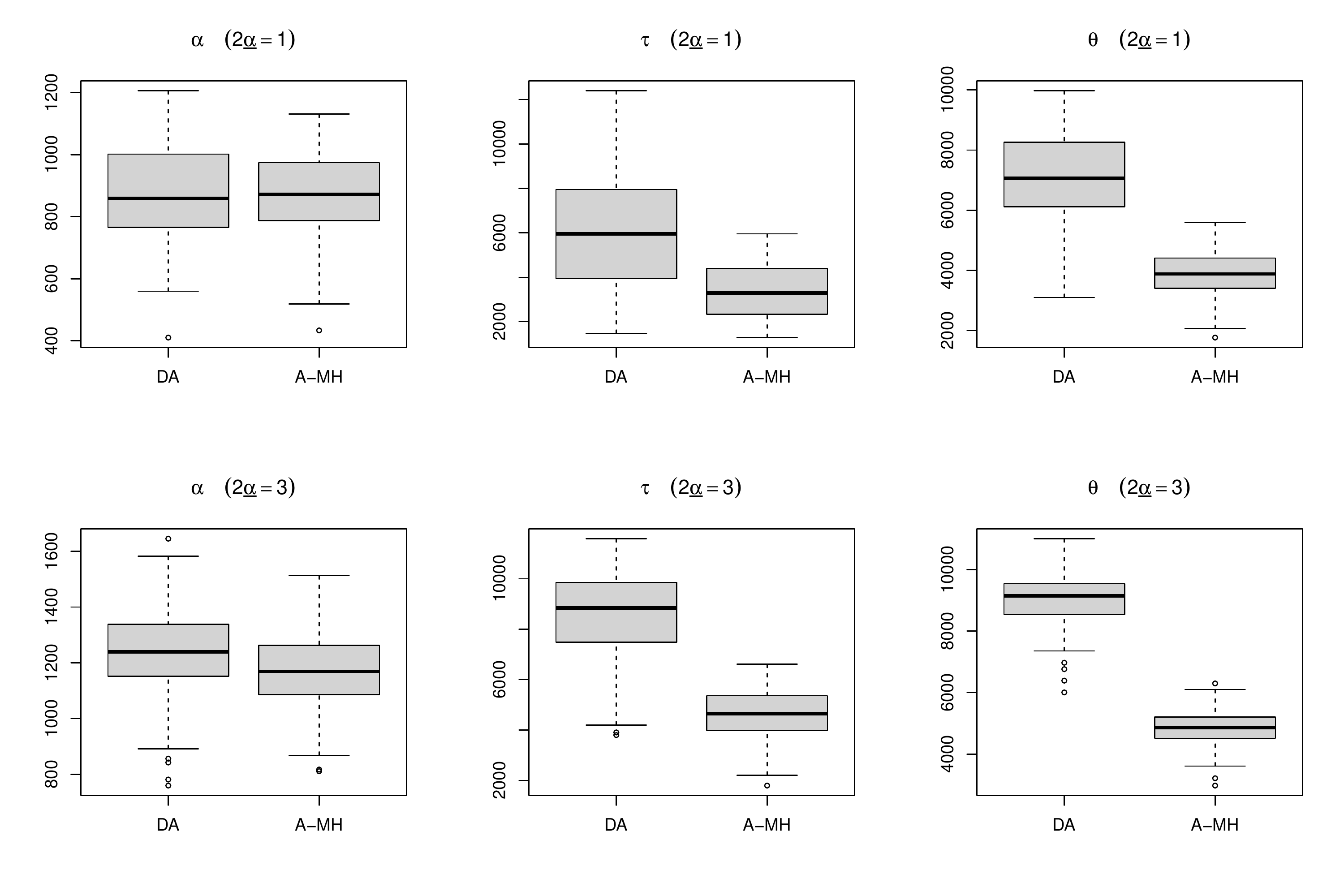}
\caption{Boxplots of the effective sample sizes standardized by the computation times for the proposed method (DA) and the alternative method (A-MH) for $2 \underline{\al } = 1, 3$ for $n = 10$. }
\label{fig:tgamma_1}
\end{figure}

Boxplots of sESSs are shown in Figure \ref{fig:tgamma_1} for $n=10$ (and in Figures~S3 and S4 for $n = 30$ and $n = 100$, respectively), and
Table~\ref{table:tgamma} lists the averages of ESSs, sESSs and the ratios of MSEs computed in this experiment. 
In these scenarios, our method becomes more competitive even for large $n$. In particular, our method outperforms the A-MH method in terms of sESS for $\th $ and $\ta $ when $n = 10$ and $n = 30$, and for $\th $ when $n = 100$.

\small
\begin{table}[!thb]
\caption{%
The averages of the effective sample sizes (ESS) %
for the proposed method (DA) and the alternative method (A-MH), the averages of those standardized by computation time (sESS), and the ratios of the mean squared errors (MSE) of the A-MH method to those of the DA method for $2 \underline{\al } = 1, 3$. }
\begin{center}
$
{\renewcommand\arraystretch{1.1}\small
\begin{array}{cccccccccccccccccccc}
\hline &&&& \multicolumn{3}{c}{\text{ESS}} && \multicolumn{3}{c}{\text{sESS}} && \multicolumn{3}{c}{\text{MSE ratio}} \\          
n & 2 \underline{\al } & \text{method} 
&& \text{$\th $} & \text{$\ta $}
& \text{$\al $} 
&& \text{$\th $} & \text{$\ta $}
& \text{$\al $}
&& \text{$\th $} & \text{$\ta $}
& \text{$\al $}
\\

\hline

10 & 	1 & \text{DA} &&		2376  & 2033	 & 293 && 7029 & 	6013  & 	867  && - & -& 	- \\

10	 & 1 & 	\text{A-MH} &&	2371 	 & 2092& 533	 && 3866 	 & 3419 & 	868  && 	1.01  & 0.98 	 & 0.96 \\

10 & 	3 & \text{DA} &&		2885 & 	2735 & 	396 && 8963	 & 8494 & 1230	 && - & -& 	- \\

10 & 	3 & \text{A-MH} &&	2893& 	2761  & 	695&& 	4877 	 & 4660  & 	1171  && 1.00 & 	1.00  & 	0.99  \\

\hline

30	 & 1	 &\text{DA} &&	 2341 	 & 601 & 	126  && 6354	 & 1627	 & 342 && 	- &- & - \\

30 & 	1	 & \text{A-MH} &&2371  & 767  & 217  && 	4074	 & 1325  & 	373 && 	1.01  & 	0.99  & 	0.89  \\

30 & 	3	 & \text{DA} &&	2710  & 1057 & 141 && 	7376 	 & 2871 	 & 383 	 && - & 	-& 	- \\

30	 & 3	 & \text{A-MH} &&2702 & 1206  & 	251 && 4567 & 	2051	 & 424  && 	1.00 & 1.03	 & 1.03  \\

\hline

100 & 	1	 & \text{DA} &&	2660 & 	247  & 59 && 	5595 	 & 521 	 & 123 	 && -& -	 & - \\

100	 & 1 & \text{A-MH} &&	2678& 346	 & 106 && 	4278 & 	552  & 168 	 && 0.99  	 & 1.00 & 	0.94  \\

100	 & 3 & 	\text{DA} &&	2679 & 	283 & 63  && 	5670	 & 600	 & 134 	 && -& 	- & - \\

100	 & 3	 & \text{A-MH} &&2719 & 381  & 	108 && 4300 & 	604 	 & 170	 && 1.00  	 & 1.03 & 	1.04  \\

\hline
\end{array}
}
$
\end{center}
\label{table:tgamma} 
\end{table}
\normalsize

\section{The Dirichlet-Multinomial Distribution}
\label{sec:dir2} 
\subsection{Sampling algorithm}
\label{subsec:dir2_algorithm} 

Dirichlet-multinomial distribution is useful for modeling multi-label variables, as used in topic modeling (e.g. Blei et al., 2003). Since the full conditional distribution of the shape parameters of the Dirichlet distribution includes the reciprocal gamma function, their posterior sampling is typically not straightforward (e.g. Nandram, 1998). 
Although in this section our focus is the estimation of the shape parameters of the Dirichlet-multinomial distribution, our result is also relevant in the context of finite mixture modeling (e.g. Fr\"{u}hwirth-Schnatter, 2006).

Suppose that for $i = 1, \dots , n$, 
\begin{align}
&\x _i \sim {\rm{Multin}}_L ( \x _i | N_i , \p _i ) = {N_i ! \over \prod_{l = 0}^{L} ( x_{i, l} !)} \prod_{l = 0}^{L} {p_{i, l}}^{x_{i, l}} \text{,} \non \\
&\p _i \sim {\rm{Dir}}_L ( \p _i | \bal ) = {\Ga \big ( \sum_{l = 0}^{L} \al _l \big) \over \prod_{l = 0}^{L} \Ga ( \al _l )} \prod_{l = 0}^{L} {p_{i, l}}^{\al _l - 1} \text{,} \non \\
&\bal \sim p( \bal ) \text{,} \non 
\end{align}
where $\x _i = ( x_{i, 1} , \dots , x_{i, L} )$, $x_{i, 0} = N_i - \sum_{l = 1}^{L} x_{i, l}$, $\p _i = ( p_{i, 1} , \dots , p_{i, L} ) \in D_L = \big\{ ( \tilde{p} _1 , \dots , \tilde{p} _L ) \in (0, 1)^L \big| \tilde{p} _1 , \dots , \tilde{p} _L > 0, \, \sum_{l = 1}^{L} \tilde{p} _l < 1 \big\} $, $p_{i, 0} = 1 - \sum_{l = 1}^{L} p_{i, l}$, and $\bal = ( \al _0 , \dots , \al _L ) $. 
Let $\x = ( \x _1 , \dots , \x _n )$ and $\p = ( \p _1 , \dots , \p _n )$. 
Since we have been unable to find good changes of variables to perform the second step of Section \ref{subsec:properties_gamma}, we use the flexible method of Section \ref{subsec:generic_method}. 
In Section 
S6 of the Supplementary Material, we prove that the posterior distribution $p( \p , \bal | \x )$ is obtained as the marginal distribution of 
\begin{align}
p( \p , \bal , \z , \w , \bro | \x ) &\propto p( \bal ) \Big\{ \prod_{l = 0}^{L} ( {\al _l}^n e^{2 n \al _l} ) \Big\} \Big( \prod_{i = 1}^{n} \prod_{l = 0}^{L} {p_{i, l}}^{x_{i, l} + \al _l - 1} \Big) \Big\{ \prod_{i = 1}^{n} ( {z_i}^{\sum_{l = 0}^{L} \al _l - 1} e^{- z_i} ) \Big\} \non \\
&\quad \times \Big\{ \prod_{l = 0}^{L} ( {w_l}^{n \al _l - 1} e^{- w_l n {\al _l}^2} ) \Big\} \Big[ \prod_{i = 2}^{n} \prod_{l = 0}^{L} \{ {\rho _{i, l}}^{\al _l + (i - 1) / n - 1} (1 - \rho _{i, l} )^{(n - i + 1) / n - 1} \} \Big] \non \\
&\quad \times \prod_{l = 0}^{L} \Big\{ {(n \al _l )^{n \al _l - 1 / 2} \over \Ga (n \al _l ) e^{n \al _l}} \Big\} ^2 \text{,} \label{eq:dir2_new_prp} 
\end{align}
where $\z = ( z_1 , \dots , z_n ) \in (0, \infty )^n$, $\w = ( w_0 , \dots , w_L ) \in (0, \infty )^{L + 1}$, and $\bro = ( \bro _2 , \dots , \bro _n ) = (( \rho _{2, 0} , \dots , \rho _{2, L} ), \dots , ( \rho _{n, 0} , \dots , \rho _{n, L} )) \in (0, 1)^{(n - 1) (L + 1)}$ are additional latent variables.

If we use the prior $p( \bal ) = \prod_{l = 0}^{L} {\rm{Ga}} ( \al _l | a, b)$ for example, we can use the following algorithm to generate posterior samples. 
\begin{algo}
\label{algo:dir2} 
The variables $\p $, $\bal $, $\z $, $\w $, and $\bro $ are updated in the following way. 
\begin{itemize}
\item[-]
Sample $\p ^{*} = (( p_{1, 0}^{*} , \dots , p_{1, L}^{*} ), \dots , ( p_{n, 0}^{*} , \dots , p_{n, L}^{*} ))\sim \prod_{i = 1}^{n} {\rm{Dir}}_L ( \x _i + \bal )$. 
\item[-]
Sample $\z ^{*} = ( z_{1}^{*} , \dots , z_{n}^{*} ) \sim \big\{ {\rm{Ga}} \big( \sum_{l = 0}^{L} \al _l , 1 \big) \big\} ^n$. 
\item[-]
Sample $\w ^{*} = ( w_{0}^{*} , \dots , w_{L}^{*} ) \sim \prod_{l = 0}^{L} {\rm{Ga}} (n \al _l , n {\al _l}^2 )$. 
\item[-]
Sample $\bro ^{*} = (( \rho _{2, 0}^{*} , \dots , \rho _{2, L}^{*} ), \dots , ( \rho _{n, 0}^{*} , \dots , \rho _{n, L}^{*} )) \sim \prod_{i = 2}^{n} \prod_{l = 0}^{L} {\rm{Beta}} ( \al _l + (i - 1) / n, (n - i + 1) / n)$. 
\item[-]
For $l = 0, \dots , L$, let $c_l = n + a $, $a_l = n w_{l}^{*}$, and 
\begin{align}
b_l = - \sum_{i = 1}^{n} \log {1 \over p_{i, l}^{*}} + \sum_{i = 1}^{n} \log z_{i}^{*} + 2 n + n \log w_{l}^{*} - \sum_{i = 2}^{n} \log {1 \over \rho _{i, l}^{*}} - b \non 
\end{align}
and sample $\al _{l}^{*}$ in one of the following three ways and accept it with probability 
\begin{align}
\min \Big\{ 1, \Big\{ {(n \al _{l}^{*} )^{n \al _{l}^{*} - 1 / 2} \over \Ga (n \al _{l}^{*} ) e^{n \al _{l}^{*}}} \Big\} ^2 / \Big\{ {(n \al _l )^{n \al _l - 1 / 2} \over \Ga (n \al _l ) e^{n \al _l}} \Big\} ^2 \Big\} \text{.} \non 
\end{align}
\begin{enumerate}
\item[{\rm{(i)}}]
Sample $\al _{l}^{*} \sim {\rm{PTN}} ( c_l , a_l , b_l )$ by using the PTN sampler developed by He et al. (2021). 
\item[{\rm{(ii)}}]
Let $M_l = 1 + \max \{ 0, b_l \} $ and ${b_l}' = M_l - b_l$. 
\begin{itemize}
\item
Sample $\ze _{l}^{*} \sim {\rm{Po}} ( M_l \al _l )$. 
\item
Sample $\eta _{l}^{*} \sim {\rm{GIG}} (1 / 2, 1, ( {b_l}' )^2 {\al _l}^2 )$. 
\item
Sample $\alt _{l}^{*} \sim {\rm{Ga}} (( \ze _{l}^{*} + c_l ) / 2, a_l + ( {b_l}' )^2 / (2 \eta _{l}^{*} ))$ and set $\al _{l}^{*} = ( \alt _{l}^{*} )^{1 / 2}$. 
\end{itemize}
\item[{\rm{(iii)}}]
Let $M_l = 1 + \max \{ 0, b_l \} $ and ${b_l}' = M_l - b_l$. 
\begin{itemize}
\item
Sample $\th _{l}^{*} \sim {\rm{N}} (2 M_l \al _l , 2 M_l \al _l )$. 
\item
Sample $\eta _{l}^{*} \sim {\rm{GIG}} (1 / 2, 1, \{ {b_l}' \al _l + {\th _l}^2 / (4 M_l \al _l ) \} ^2 )$. 
\item
Sample $\alt _{l}^{*} \sim {\rm{GIG}} ( c_l / 2 - 1 / 4, 2 a_l + ( {b_l}' )^2 / \eta _{l}^{*} , ( \th _{l}^{*} )^4 / (16 {M_l}^2 \eta _{l}^{*} ))$ and set $\al _{l}^{*} = ( \alt _{l}^{*} )^{1 / 2}$. 
\end{itemize}
\end{enumerate}
\end{itemize}
\end{algo}

\subsection{Simulation study}
\label{subsec:dir2_sim} 
In this section, we conduct a simulation study-- the posterior inference of Dirichlet shape parameters-- to compare our method and the method of He et al. (2021). Both methods are based on data augmentation but in different ways. Many other standard methods, including one by Miller (2019), are not directly applicable. 

Following He et al. (2021), we set $L + 1 = 10$ and $N_1 = \dots = N_n = 500$ and use the prior $p( \bal ) = \prod_{j = 0}^{9} {\rm{Ga}} ( \al _l | b / 10, b)$ with $b = 1$. 
We generate $\p _i$ from ${\rm{Dir}}_9 ( \bal _0 )$ and then $\x _i$ from ${\rm{Multin}}_9 (500, \p _i )$. 
We consider the cases $n = 100$ and $n = 1000$. 
For each of these cases, we consider two scenarios: (I) $\bal _0 = (1 / 10, \dots , 1 / 10)$ (equal case) and (II) $\bal _0 = (1 / 10, 2 / 10, \dots , 10 / 10)$. 
Other scenarios are also considered and reported in the Supplementary Material. 
We generate $4,000$ posterior samples after discarding the first $1,000$ samples. 
We repeat this simulation $100$ times. 
The method of He et al. (2021) requires sampling from the exponential reciprocal gamma (ERG) distribution, for which they gave three methods. 
We use the first method because it is the easiest to implement. 
Setting $N$ equal to a large value in (16) of He et al. (2021) makes their approximation accurate. 
We set $N = 3$, so that their approximation is sufficiently accurate. %

We consider the proposed method based on (iii), (ii) and (i) of Algorithm \ref{algo:dir2} (DA-N, DA-P and DA-PT, respectively), as well as the method of He et al. (2021) (ERG). 
Using these methods, we calculate the averages over the simulations of the means of the effective sample sizes %
for $\al _0 , \dots , \al _9$ (ESS), the averages over the simulations of the computation times (CT), and the averages over the simulations of the ratios of the means of the effective sample sizes to the computation times (sESS). 
We also calculate the mean squared errors (MSE) of the estimators of $\al _0 , \dots , \al _9$.

The results are reported in Table \ref{table:dir2}. 
In all scenarios, the ERG method has the largest ESS but the longest CT. 
In contrast, the DA-N, DA-P and DA-PT methods are less competitive than the ERG method in ESSs, but significantly outperform it in computational time. Consequently, all of our methods have much larger sESSs than the state-of-the-art ERG method. 
Among the proposed methods, the DA-PT method has the best sESS. The other two methods cost computational efficiency for the simplicity of their algorithms, as noted in Section~\ref{subsec:generic_method}. In terms of MSE, no significant difference can be seen in the four methods.

\small
\begin{table}[!thb]
\caption{
The average effective sample size (ESS), the average computation time (CT), the standardized effective sample size by the computation time (sESS), and the mean squared error (MSE) for the proposed data-augmentation method with normal latent variables (DA-N), Poisson latent variables (DA-P), and the PTN sampler of He et al. (2021) (DA-PT) and the original method proposed by He et al. (2021) (ERG). 
These values are averaged over $\al _0 , \dots , \al _9$. 
MSE values under $n=100$ and $n=1000$ are multiplied by $10^3$ and $10^4$, respectively. 
}
\begin{center}
$
{\renewcommand\arraystretch{1.1}\small
\begin{array}{ccccccccccccccccccc}
\hline
n & \text{Scenario} & \text{method} && \text{ESS} & \text{CT} & \text{sESS} & \text{MSE} \\

\hline

100	 & \text{(I)}	 & \text{DA-N} && 863 & 	2.1 & 414 & 	0.82  \\

 &  	 & \text{DA-P} && 856	 & 1.8 	 & 480	 & 0.83 \\

& 	  & \text{DA-PT} && 	1199 	 & 1.8& 667& 	0.82  \\

 &  & \text{ERG} &&	1808  & 	81.4& 	22  & 0.82 \\

\hline

100	 & \text{(II)} & \text{DA-N} && 	580 	 & 2.0	 & 287 	 & 5.70 \\

& 	  & \text{DA-P} && 	668 & 1.7  & 	389 & 5.68 \\

	 &   & 	\text{DA-PT} &&  846 & 	1.8  & 478  & 	 5.73 \\

 & 	 & \text{ERG} &&	1315 & 79.5 & 	17 & 	 5.66  \\

\hline

1000	 & \text{(I)} & \text{DA-N} && 	846 & 	7.7 & 	110	 & 0.80  \\

 & 	 	 &\text{DA-P} &&  846& 	7.4 & 	115  & 0.80  \\

 & 	 & \text{DA-PT} &&	1192 	 & 7.4 & 	162  & 0.80 \\

&   & \text{ERG} &&	1800 & 	819.1  & 	2  & 0.80  \\

\hline

1000 & 	\text{(II)} & \text{DA-N} && 	582	 & 7.6  & 	76  & 	 5.06 \\

 & 	  & \text{DA-P} && 	671  & 7.2 & 	93& 	 5.18 \\

 & 	  & \text{DA-PT} &&	834  & 	7.3& 	115 & 	 5.15  \\

& 	  & \text{ERG} &&	1317  & 	835.4 & 	2 & 	 5.17 \\

\hline
\end{array}
}
$
\end{center}
\label{table:dir2} 
\end{table}
\normalsize

Thus, the difference of the method of He et al. (2021) and ours in computational efficiency critically depends on the computational time. For the fairness of comparison, it should be noted that the computation by the ERG method can speed-up by using parallelization, and could be competitive as our methods in some computational environments that enable such parallelization. Other than the efficiency, the advantage of our method to be emphasized is its simplicity; no explicit parallelization is needed in implementing our method. In addition, our method is tuning parameter free, while the ERG method requires tuning $N$.

\section{Concluding Remarks}
\label{sec:conclusion}
The data augmentation approach proposed in this paper can be applicable to any posterior inference if the conditional posterior involves the reciprocal gamma functions. 
Examples of such models include the one-parameter Dirichlet, negative binomial and Wishart models, in addition to the gamma, Student's $t$ and Dirichlet-multinomial models considered in the previous sections. 
The sampling algorithms for those models can be derived straightforwardly and are provided in Section S1 of the Supplementary Material.

A remaining issue related to the proposed approach is that our method is likely to be less efficient for extremely small $n$. 
In that case, the data augmentation in Theorem~
S1 should be customized for the model of interest. 
For example, if $n = 1$ and $0 < \al _l \ll 1$ in the Dirichlet-multinomial model, we could improve the proposed augmentation; see Section 
S5 of the Supplementary Material.

\section*{Acknowledgments}
Research of the authors was supported in part by JSPS KAKENHI Grant Number 20J10427, 19K11852, 17K17659, and 21H00699 from Japan Society for the Promotion of Science.

\newpage
\setcounter{equation}{0}
\renewcommand{\theequation}{S\arabic{equation}}
\setcounter{section}{0}
\renewcommand{\thesection}{S\arabic{section}}
\setcounter{lem}{0}
\renewcommand{\thelem}{S\arabic{lem}}
\setcounter{thm}{0}
\renewcommand{\thethm}{S\arabic{thm}}
\setcounter{table}{0}
\renewcommand{\thetable}{S\arabic{table}}
\setcounter{figure}{0}
\renewcommand{\thetable}{S\arabic{figure}}

\renewcommand{\thefigure}{S\arabic{figure}}
\renewcommand{\thetable}{S\arabic{table}}

\begin{center}
{\LARGE {\bf Supplementary Material for ``On Data Augmentation for Models Involving Reciprocal Gamma Functions"}}
\end{center}

\vspace{1cm}
In Section \ref{subsec:other_models}, we briefly discuss how we could use our method for three models not considered in the main text. 
In Section \ref{sec:proof_thm_1}, we prove Theorem 
1 of the main text. 
In Section \ref{subsec:lemmas}, we prove two lemmas. 
In Section \ref{subsec:proof_cor}, we prove Theorem \ref{thm:general}, from which Theorem 
1 and the theorems of Section \ref{subsec:extensions} follow. 
In Section \ref{subsec:extensions}, we consider extensions of Theorem 
1. 
In Section \ref{subsec:proof_prp}, we derive the expressions 
(3) and (4) of the main text and the expressions (\ref{eq:dir1_prp}), (\ref{eq:nb_prp}), (\ref{eq:wishart_prp_1}), and (\ref{eq:wishart_prp_2}) of this Supplementary Material. 
In Section \ref{subsec:miller_t}, details of the alternative method considered in Section 
3.2 of the main text are given. 
Additional results for the simulation study of Section 
3.2 of the main text are in Section \ref{subsec:additional_figures}.

\section{Other examples}
\label{subsec:other_models} 
Here, we consider three additional models to which our approach is relevant. 
In Sections \ref{subsubsec:dir1}, \ref{subsubsec:nb}, and \ref{subsubsec:wishart}, we consider models based on the one-parameter Dirichlet distribution, the negative binomial distribution, and the Wishart distribution, respectively. 
The expressions (\ref{eq:dir1_prp}), (\ref{eq:nb_prp}), (\ref{eq:wishart_prp_1}), and (\ref{eq:wishart_prp_2}) are derived in Section \ref{subsec:proof_prp} of this Supplementary Material.

\subsection{The one-parameter Dirichlet prior distribution}
\label{subsubsec:dir1} 
Here, we first consider a simple model based on the Dirichlet distribution with a single shape parameter. 
Suppose that for $i = 1, \dots , n$, 
\begin{align}
&\x _i \sim p_i ( \x _i | \p _i ) \text{,} \non \\
&\p _i \sim {\rm{Dir}}_L ( \p _i | \al , \dots , \al ) = {\Ga ((L + 1) \al ) \over \{ \Ga ( \al ) \} ^{L + 1}} \prod_{l = 0}^{L} {p_{i, l}}^{\al - 1} \text{,} \non \\
&\al \sim p( \al ) \text{,} \non 
\end{align}
where $\x _i \in \mathcal{X} _i$, $\p _i = ( p_{i, 1} , \dots , p_{i, L} ) \in D_L = \big\{ ( \tilde{p} _1 , \dots , \tilde{p} _L ) \in (0, 1)^L \big| \tilde{p} _1 , \dots , \tilde{p} _L > 0, \, \sum_{l = 1}^{L} \tilde{p} _l < 1 \big\} $, and $p_{i, 0} = 1 - \sum_{l = 1}^{L} p_{i, l}$. 
Then the posterior distribution $p( \p , \al | \x )$ is obtained as the marginal distribution of 
\begin{align}
p( \p , \al , \bro | \x ) &\propto p( \al ) \al ^{n L} \exp \Big( - \al \sum_{i = 1}^{n} \Big[ \sum_{l = 1}^{L} \log {1 \over \rho _{i, l}} + \sum_{l = 0}^{L} \Big\{ \log {1 \over p_{i, l}} - \log (L + 1) \Big\} \Big] \Big) \non \\
&\quad \times \Big[ \prod_{i = 1}^{n} \prod_{l = 1}^{L} \{ {\rho _{i, l}}^{l / (L + 1) - 1} (1 - \rho _{i, l} )^{(L - l + 1) / (L + 1) - 1} \} \Big] \Big( \prod _{i = 1}^{n} \prod_{l = 0}^{L} {p_{i, l}}^{- 1} \Big) \prod _{i = 1}^{n} p_i ( \x _i | \p _i ) \text{,} \label{eq:dir1_prp} 
\end{align}
where $\bro = ( \bro _1 , \dots , \bro _n ) = (( \rho _{1, 1} , \dots , \rho _{1, L} ), \dots , ( \rho _{n, 1} , \dots , \rho _{n, L} )) \in (0, 1)^{n L}$ is a set of additional latent variables.

By (\ref{eq:dir1_prp}), we can construct a Gibbs sampler without a MH step, for example, if $p( \x _i ) = {\rm{Multin}}_L ( \x _i | N_i , \p _i ) = \big\{ ( N_i !) / \prod_{l = 0}^{L} ( x_{i, l} !) \big\} \prod_{l = 0}^{L} {p_{i, l}}^{x_{i, l}} $ for $i = 1, \dots , n$ and $p( \al ) = {\rm{Ga}} ( \al | a, b)$ for $a, b > 0$, where $( x_{i, 1} , \dots , x_{i, L} ) = \x _i$ and $x_{i, 0} = N_i - \sum_{l = 1}^{L} x_{i, l}$ for $i = 1, \dots , n$. 
The full conditional distributions are as follows. 
\begin{itemize}
\item
The full conditional distributions of $\p _i$ are ${\rm{Dir}}_L ( \p _i | x_{i, 0} + \al , \dots , x_{i, L} + \al )$. 
\item
The full conditional distributions of $\rho _{i, l}$ are ${\rm{Beta}} ( \rho _{i, l} | \al + l / (L + 1), (L - l + 1) / (L + 1))$. 
\item
The full conditional distribution of $\al $ is 
\begin{align}
&{\rm{Ga}} \Big( \al \Big| n L + a, \sum_{i = 1}^{n} \Big[ \sum_{l = 1}^{L} \log {1 \over \rho _{i, l}} + \sum_{l = 0}^{L} \Big\{ \log {1 \over p_{i, l}} - \log (L + 1) \Big\} \Big] + b \Big) \text{,} \non 
\end{align}
where for all $i = 1, \dots , n$, we have $\sum_{l = 0}^{L} \{ \log (1 / p_{i, l} ) - \log (L + 1) \} > 0$ by Jensen's inequality. 
\end{itemize}

\subsection{The negative binomial distribution} 
\label{subsubsec:nb} 
Here, we consider the estimation of a negative binomial shape parameter. 
Suppose that for $i = 1, \dots , n$, 
\begin{align}
&y_i \sim {\rm{NB}} ( y_i | \al , p_i ) = {\Ga ( \al + y_i ) \over y_i ! \Ga ( \al )} {p_i}^{\al } (1 - p_i )^{y_i} \text{,} \non \\
&\al \sim p( \al ) \text{.} \non 
\end{align}
We assume that $p_1 , \dots , p_n \in (0, 1)$ are known and fixed for simplicity. 
However, we can consider the famous negative binomial regression model by setting $p_i = {\x _i}^{\top } \bbe $ for known $\x _i \in \mathbb{R} ^p$ and unknown $\bbe \sim \pi ( \bbe )$ and using the data augmentation scheme of Polson et al. (2013); see He et al. (2021). 

Let $\y = ( y_1 , \dots , y_n )$. 
The posterior distribution $p( \al | \y )$ is obtained as the marginal distribution of 
\begin{align}
p( \al , \z , w, \bro | \y ) &\propto p( \al ) \al ^n e^{2 n \al } \Big( \prod_{i = 1}^{n} p_i \Big) ^{\al } \Big\{ \prod_{i = 1}^{n} ( {z_i}^{\al + y_i - 1} e^{- z_i} ) \Big\} w^{n \al - 1} e^{- w n {\al }^2} \non \\
&\quad \times \Big[ \prod_{i = 2}^{n} \{ {\rho _i}^{\al + (i - 1) / n - 1} (1 - \rho _i )^{(n - i + 1) / n - 1} \} \Big] \Big\{ {(n \al )^{n \al - 1 / 2} \over \Ga (n \al ) e^{n \al }} \Big\} ^2 \text{,} \label{eq:nb_prp} 
\end{align}
where $\z = ( z_1 , \dots , z_n ) \in (0, \infty )^n$, $w \in (0, \infty )$, and $\bro = ( \rho _2 , \dots , \rho _n ) \in (0, 1)^{n - 1}$ are additional latent variables. 
The approaches outlined in Section 
2.4 could be useful when we use (\ref{eq:nb_prp}). 
An alternative approach would be to combine Theorem 
1 with the result of Zhou and Carin (2015).

\subsection{The Wishart distribution} 
\label{subsubsec:wishart} 
Estimating the shape parameter of the Wishart distribution is also an important problem. 
For example, Xiao et al. (2015) discussed the importance of considering a fractional shape parameter in the context of analyzing time series data. 

Consider the simple model where for $i = 1, \dots , n$, 
\begin{align}
&\x _i \sim {\rm{N}}_p ( \x _i | \bm{0} _p , \Psi ^{- 1} ) = {1 \over (2 \pi )^{p / 2}} | \bPsi |^{1 / 2} e^{- {\x _i}^{\top } \bPsi \x _i / 2} \text{,} \non \\
&\bPsi \sim {\rm{W}}_p ( \bPsi | 2 \al + p - 1, ( \be \I _p )^{- 1} ) = {\be ^{p \{ \al + (p - 1) / 2 \} } \over 2^{p \{ \al + (p - 1) / 2 \} } \Ga _p ( \al + (p - 1) / 2)} | \bPsi |^{\al - 1} e^{- \be \tr \bPsi / 2} \text{,} \non \\
&( \al , \be ) \sim {\rm{Ga}} ( \al | a, b) {\rm{Ga}} ( \be | c, d) g( \al , \be ) \text{.} \non 
\end{align}
In this case, although Theorem 
1 is not applicable to $1 / \Ga _p ( \al + (p - 1) / 2)$, we can derive a convenient expression based on a similar idea; see Lemma \ref{lem:wishart} of Section \ref{subsec:proof_prp}. 
Let $\ga = \be / \al $. 
Our results are the following. 
\begin{itemize}
\item[{\rm{(i)}}]
If $p = 2 m$ for $m \in \mathbb{N}$, then the posterior distribution $p( \bPsi , \al , \ga | \x )$ is obtained as the marginal distribution of 
\begin{align}
p( \bPsi , \al , \ga , \bro | \x ) &\propto g( \al , \al \ga ) {e^{2 m \al } \over {\al }^{- 1 / 2}} {\al ^{p (p - 1) / 2 + c + a - 1} e^{- b \al } \ga ^{p \{ \al + (p - 1) / 2 \} + c - 1} e^{- d \al \ga } \over 2^{p \al }} \non \\
&\quad \times | \bPsi |^{n / 2 + \al - 1} e^{- \tr ( \al \ga \I _p + \sum_{i = 1}^{n} \x _i {\x _i}^{\top } ) \bPsi / 2} \non \\
&\quad \times \Big[ \prod_{j = 2}^{m} \{ {\rho _j}^{2 \al + (j - 1) / m - 1} (1 - \rho _j )^{(2 - 1 / m) (j - 1) - 1} \} \Big] {(2 m \al )^{2 m \al - 1 / 2} \over \Ga (2 m \al ) e^{2 m \al }} \text{.} \label{eq:wishart_prp_1} 
\end{align}
where $\bro = ( \rho _2 , \dots , \rho _m ) \in (0, 1)^{m - 1}$ is a set of additional latent variables. 
\item[{\rm{(ii)}}]
If $p = 2 m - 1$ for $m \in \mathbb{N}$, then the posterior distribution $p( \bPsi , \al , \ga | \x )$ is obtained as the marginal distribution of 
\begin{align}
p( \bPsi , \al , \ga , \bro , z | \x ) &\propto g( \al , \al \ga ) {e^{2 m \al } \over {\al }^{1 / 2}} {\al ^{p (p - 1) / 2 + c + a - 1} e^{- b \al } \ga ^{p \{ \al + (p - 1) / 2 \} + c - 1} e^{- d \al \ga } \over 2^{p \al }} \non \\
&\quad \times | \bPsi |^{n / 2 + \al - 1} e^{- \tr ( \al \ga \I _p + \sum_{i = 1}^{n} \x _i {\x _i}^{\top } ) \bPsi / 2} \non \\
&\quad \times \Big[ \prod_{j = 2}^{m} \{ {\rho _j}^{2 \al + (j - 1) / m - 1} (1 - \rho _j )^{(2 - 1 / m) (j - 1) - 1} \} \Big] z^{\al + m - 1 / 2 - 1} e^{- z} {(2 m \al )^{2 m \al - 1 / 2} \over \Ga (2 m \al ) e^{2 m \al }} \text{,} \label{eq:wishart_prp_2} 
\end{align}
where $\bro $ and $z \in (0, \infty )$ are additional latent variables. 
\end{itemize}

For simplicity, here we consider only the case of even $p$. 
Let $m = p / 2 \in \mathbb{N}$. 
The full conditional distributions under the gamma prior $p( \al , \be ) = {\rm{Ga}} ( \al | a, b) {\rm{Ga}} ( \be | c, d)$ are as follows. 
\begin{itemize}
\item
The full conditional of $\bPsi $ is ${\rm{W}}_p \big( \bPsi \big| n + 2 \al + p - 1, \big( \al \ga \I _p + \sum_{i = 1}^{n} \x _i {\x _i}^{\top } \big) ^{- 1} \big) $. 
\item
The full conditional distributions of $\rho _j$ are ${\rm{Beta}} ( \rho _j | 2 \al + (j - 1) / m, (2 - 1 / m) (j - 1))$. 
\item
The full conditional of $\ga $ is ${\rm{Ga}} ( \ga | p \{ \al + (p - 1) / 2 \} + c, ( \tr \bPsi / 2 + d) \al )$. 
\item
The full conditional of $\al $ is 
\begin{align}
p( \al | \bPsi , \ga , \bro , \x ) &\propto {\rm{Ga}} \Big( \al \Big| {1 \over 2} + {p (p - 1) \over 2} + c + a, B \Big) {(p \al )^{p \al - 1 / 2} \over \Ga (p \al ) e^{p \al }} \text{,} \non 
\end{align}
where \begin{align}
B &= b + d \ga + {\ga \over 2} \tr \bPsi - p - \Big( p \log {\ga \over 2} + \log | \bPsi | \Big) + \sum_{j = 2}^{m} 2 \log {1 \over \rho _j} \non 
\end{align}
is positive since the full conditional must be proper and since the factor $(p \al )^{p \al - 1 / 2} / \{ \Ga (p \al ) e^{p \al } \} $ is bounded as in 
(1). 
\end{itemize}

\section{Proof of Theorem 
1}
\label{sec:proof_thm_1} 
Here, we prove Theorem 
1 of the main text. 

\bigskip

\noindent
{\bf Proof of Theorem 
1.} \ \ We first write $1 / \{ \Ga ( \xi ) \} ^m$ as 
\begin{align}
{1 \over \{ \Ga ( \xi ) \} ^m} &= {1 \over \prod_{j = 0}^{m - 1} \Ga ( \xi + j / m)} \prod_{j = 1}^{m} {\Ga ( \xi + (j - 1) / m) \over \Ga ( \xi )} \text{.} \non 
\end{align}
By Gauss's multiplication formula, we have 
\begin{align}
{1 \over \prod_{j = 0}^{m - 1} \Ga ( \xi + j / m)} &= {(2 \pi )^{(1 - m) / 2} m^{m \xi - 1 / 2} \over \Ga (m \xi )} \text{.} \non 
\end{align}
On the other hand, for all $j = 1, \dots , m$, 
\begin{align}
{\Ga ( \xi + (j - 1) / m) \over \Ga ( \xi )} &= {\Ga ( \xi + (j - 1) / m) \over \Ga ( \xi )} \int_{0}^{1} {\rm{Beta}} \Big( \rho _j \Big| \xi + {j - 1 \over m}, {m - j + 1 \over m} \Big) d{\rho _j} \non \\
&= {\xi \over \Ga ((m - j + 1) / m)} \int_{0}^{1} {\rho _j}^{\xi + (j - 1) / m - 1} (1 - \rho _j )^{(m - j + 1) / m - 1} d{\rho _j} \text{.} \non 
\end{align}
Therefore, 
\begin{align}
{1 \over \{ \Ga ( \xi ) \} ^m} &= {(2 \pi )^{(1 - m) / 2} m^{m \xi - 1 / 2} \over \Ga (m \xi )} \prod_{j = 2}^{m} {\xi \over \Ga ((m - j + 1) / m)} \int_{0}^{1} {\rho _j}^{\xi + (j - 1) / m - 1} (1 - \rho _j )^{(m - j + 1) / m - 1} d{\rho _j} \text{,} \non 
\end{align}
and the result follows. 
\hfill$\Box$

\section{Lemmas}
\label{subsec:lemmas} 
Here, we present two lemmas.

\begin{lem}
\label{lem:2} 
Let $M \in \mathbb{N} _0$. 
Then 
\begin{align}
{1 \over \Ga ( \xi )} &= {(M + 1)^{- 1 / 2} \over (2 \pi )^{M / 2}} {(M + 1)^{(M + 1) \xi } \over \Ga ((M + 1) \xi )} \prod_{j = 2}^{M + 1} \int_{0}^{\infty } {t_j}^{\xi + (j - 1) / (M + 1) - 1} e^{- t_j} d{t_j} \non 
\end{align}
for all $\xi > 0$. 
\end{lem}

\noindent
{\bf Proof.} \ \ By Gauss's multiplication formula, 
\begin{align}
{1 \over \Ga ( \xi )} &= {1 \over \prod_{j = 0}^{M} \Ga ( \xi + j / (M + 1))} \prod_{j = 2}^{M + 1} \Ga ( \xi + (j - 1) / (M + 1)) \non \\
&= {(2 \pi )^{- M / 2} (M + 1)^{(M + 1) \xi - 1 / 2} \over \Ga ((M + 1) \xi )} \prod_{j = 2}^{M + 1} \int_{0}^{\infty } {t_j}^{\xi + (j - 1) / (M + 1) - 1} e^{- t_j} d{t_j} \text{,} \non 
\end{align}
which is the desired result. 
\hfill$\Box$

\begin{lem}
\label{lem:3} 
Let $K \in \mathbb{N} _0$. 
Let $M_1 , \dots , M_K \in \mathbb{N} _0$. 
Then 
\begin{align}
{1 \over \Ga ( \xi )} &= \Big\{ \prod_{k = 1}^{K} {( M_k + 1)^{- 1 / 2} \over (2 \pi )^{M_k / 2}} \Big\} {\prod_{k = 1}^{K} ( M_k + 1)^{( M_k + 1) ( M_{k - 1} + 1) \dotsm ( M_1 + 1) \xi } \over \Ga (( M_K + 1) \dotsm ( M_1 + 1) \xi )} \non \\
&\quad \times \prod_{k = 1}^{K} \prod_{j = 2}^{M_k + 1} \int_{0}^{\infty } {t_{k, j}}^{( M_{k - 1} + 1) \dotsm ( M_1 + 1) \xi + (j - 1) / ( M_k + 1) - 1} e^{- t_{k, j}} d{t_{k, j}} \non 
\end{align}
for all $\xi > 0$. 
\end{lem}

\noindent
{\bf Proof.} \ \ By Lemma \ref{lem:2}, we have 
\begin{align}
&{1 \over \Ga ( \xi )} = {( M_1 + 1)^{- 1 / 2} \over (2 \pi )^{M_1 / 2}} {( M_1 + 1)^{( M_1 + 1) \xi } \over \Ga (( M_1 + 1) \xi )} \prod_{j = 2}^{M_1 + 1} \int_{0}^{\infty } {t_{1, j}}^{\xi + (j - 1) / ( M_1 + 1) - 1} e^{- t_{1, j}} d{t_{1, j}} \text{,} \non \\
&{1 \over \Ga (( M_1 + 1) \xi )} = {( M_2 + 1)^{- 1 / 2} \over (2 \pi )^{M_2 / 2}} {( M_2 + 1)^{( M_2 + 1) ( M_1 + 1) \xi } \over \Ga (( M_2 + 1) ( M_1 + 1) \xi )} \prod_{j = 2}^{M_2 + 1} \int_{0}^{\infty } {t_{2, j}}^{( M_1 + 1) \xi + (j - 1) / ( M_2 + 1) - 1} e^{- t_{2, j}} d{t_{2, j}} \text{,} \non \\
&\quad \vdots \non \\
&{1 \over \Ga (( M_{K - 1} + 1) \dotsm ( M_1 + 1) \xi )} = {( M_K + 1)^{- 1 / 2} \over (2 \pi )^{M_K / 2}} {( M_K + 1)^{( M_K + 1) ( M_{K - 1} + 1) \dotsm ( M_1 + 1) \xi } \over \Ga (( M_K + 1) ( M_{K - 1} + 1) \dotsm ( M_1 + 1) \xi )} \non \\
&\quad \times \prod_{j = 2}^{M_K + 1} \int_{0}^{\infty } {t_{K, j}}^{( M_{K - 1} + 1) \dotsm ( M_1 + 1) \xi + (j - 1) / ( M_K + 1) - 1} e^{- t_{K, j}} d{t_{K, j}} \text{.} \non 
\end{align}
Therefore, 
\begin{align}
{1 \over \Ga ( \xi )} &= \Big\{ \prod_{k = 1}^{K} {( M_k + 1)^{- 1 / 2} \over (2 \pi )^{M_k / 2}} \Big\} {\prod_{k = 1}^{K} ( M_k + 1)^{( M_k + 1) ( M_{k - 1} + 1) \dotsm ( M_1 + 1) \xi } \over \Ga (( M_K + 1) \dotsm ( M_1 + 1) \xi )} \non \\
&\quad \times \prod_{k = 1}^{K} \prod_{j = 2}^{M_k + 1} \int_{0}^{\infty } {t_{k, j}}^{( M_{k - 1} + 1) \dotsm ( M_1 + 1) \xi + (j - 1) / ( M_k + 1) - 1} e^{- t_{k, j}} d{t_{k, j}} \text{,} \non 
\end{align}
which is the desired result. 
\hfill$\Box$

\section{General explicit expressions for $1 / \{ \Ga ( \xi ) \} ^m$ for beta-gamma data augmentation}
\label{subsec:proof_cor} 
Theorem 
1 follows from part (i) of the following theorem. 

\begin{thm}
\label{thm:general} 
Let $m \in \mathbb{N}$. 
Let $K \in \mathbb{N} _0$ and let $M_1 , \dots , M_K \in \mathbb{N} _0$. 
\begin{itemize}
\item[{\rm{(i)}}]
For all $\xi > 0$, we have 
\begin{align}
{1 \over \{ \Ga ( \xi ) \} ^m} &= {\{ ( M_K + 1) \dotsm ( M_1 + 1) m / (2 \pi ) \} ^{\sum_{k = 1}^{K} M_k / 2} \over (2 \pi )^{(m - 1) / 2} \prod_{j = 2}^{m} \Ga ((m - j + 1) / m)} {\xi ^{m + \sum_{k = 1}^{K} M_k / 2 + 1 / 2 - 1} e^{( M_K + 1) \dotsm ( M_1 + 1) m \xi } \over \{ ( M_K + 1) \dotsm ( M_1 + 1) \} ^{m \xi } \xi ^{m \xi }} \non \\
&\quad \times \Big\{ \prod_{k = 1}^{K} ( M_k + 1)^{( M_k + 1) \dotsm ( M_1 + 1) m \xi } \Big\} \Big\{ \prod_{j = 2}^{m} \int_{0}^{1} {\rho _j}^{\xi + (j - 1) / m - 1} (1 - \rho _j )^{(m - j + 1) / m - 1} d{\rho _j} \Big\} \non \\
&\quad \times \Big\{ \prod_{k = 1}^{K} \prod_{j = 2}^{M_k + 1} \int_{0}^{\infty } {t_{k, j}}^{( M_{k - 1} + 1) \dotsm ( M_1 + 1) m \xi + (j - 1) / ( M_k + 1) - 1} e^{- t_{k, j} ( M_K + 1) \dotsm ( M_1 + 1) m \xi } d{t_{k, j}} \Big\} \non \\
&\quad \times {\{ ( M_K + 1) \dotsm ( M_1 + 1) m \xi \} ^{( M_K + 1) \dotsm ( M_1 + 1) m \xi - 1 / 2} \over \Ga (( M_K + 1) \dotsm ( M_1 + 1) m \xi ) e^{( M_K + 1) \dotsm ( M_1 + 1) m \xi }} \text{.} \non 
\end{align}
\item[{\rm{(ii)}}]
For all $\xi > 0$, we have 
\begin{align}
&{1 \over \{ \Ga ( \xi ) \} ^m} \non \\
&= {\{ ( M_K + 1) \dotsm ( M_1 + 1) m / (2 \pi ) \} ^{\sum_{k = 1}^{K} M_k / 2} \over (2 \pi )^{(m - 1) / 2} \prod_{j = 2}^{m} \Ga ((m - j + 1) / m)} \xi ^m {( \xi + 1)^{\sum_{k = 1}^{K} M_k / 2 + 1 / 2 - 1} e^{( M_K + 1) \dotsm ( M_1 + 1) m ( \xi + 1)} \over \{ ( M_K + 1) \dotsm ( M_1 + 1) \} ^{m ( \xi + 1)} ( \xi + 1)^{m \xi }} \non \\
&\quad \times \Big\{ \prod_{k = 1}^{K} ( M_k + 1)^{( M_k + 1) \dotsm ( M_1 + 1) m ( \xi + 1)} \Big\} \Big\{ \prod_{j = 2}^{m} \int_{0}^{1} {\rho _j}^{\xi + 1 + (j - 1) / m - 1} (1 - \rho _j )^{(m - j + 1) / m - 1} d{\rho _j} \Big\} \non \\
&\quad \times \Big\{ \prod_{k = 1}^{K} \prod_{j = 2}^{M_k + 1} \int_{0}^{\infty } {t_{k, j}}^{( M_{k - 1} + 1) \dotsm ( M_1 + 1) m ( \xi + 1) + (j - 1) / ( M_k + 1) - 1} e^{- t_{k, j} ( M_K + 1) \dotsm ( M_1 + 1) m ( \xi + 1)} d{t_{k, j}} \Big\} \non \\
&\quad \times {\{ ( M_K + 1) \dotsm ( M_1 + 1) m ( \xi + 1) \} ^{( M_K + 1) \dotsm ( M_1 + 1) m ( \xi + 1) - 1 / 2} \over \Ga (( M_K + 1) \dotsm ( M_1 + 1) m ( \xi + 1)) e^{( M_K + 1) \dotsm ( M_1 + 1) m ( \xi + 1)}} \text{.} \non 
\end{align}
\end{itemize}
\end{thm}

\noindent
{\bf Proof.} \ \ Part (ii) follows from part (i). 
For part (i), we have, by Theorem 
1, 
\begin{align}
{1 \over \{ \Ga ( \xi ) \} ^m} &= {(2 \pi )^{(1 - m) / 2} m^{- 1 / 2} \over \prod_{j = 2}^{m} \Ga ((m - j + 1) / m)} {\xi ^{m - 1} m^{m \xi } \over \Ga (m \xi )} \prod_{j = 2}^{m} \int_{0}^{1} {\rho _j}^{\xi + (j - 1) / m - 1} (1 - \rho _j )^{(m - j + 1) / m - 1} d{\rho _j} \text{.} \non 
\end{align}
By Lemma \ref{lem:3}, 
\begin{align}
{1 \over \Ga (m \xi )} &= \Big\{ \prod_{k = 1}^{K} {( M_k + 1)^{- 1 / 2} \over (2 \pi )^{M_k / 2}} \Big\} {\prod_{k = 1}^{K} ( M_k + 1)^{( M_k + 1) ( M_{k - 1} + 1) \dotsm ( M_1 + 1) m \xi } \over \Ga (( M_K + 1) \dotsm ( M_1 + 1) m \xi )} \non \\
&\quad \times \prod_{k = 1}^{K} \prod_{j = 2}^{M_k + 1} \int_{0}^{\infty } {t_{k, j}}^{( M_{k - 1} + 1) \dotsm ( M_1 + 1) m \xi + (j - 1) / ( M_k + 1) - 1} e^{- t_{k, j}} d{t_{k, j}} \text{.} \non 
\end{align}
By making the change of variables $t_{k, j} = \tilde{t} _{k, j} ( M_K + 1) \dotsm ( M_1 + 1) m \xi $, 
\begin{align}
{1 \over \Ga (m \xi )} &= \Big\{ \prod_{k = 1}^{K} {( M_k + 1)^{- 1 / 2} \over (2 \pi )^{M_k / 2}} \Big\} {\prod_{k = 1}^{K} ( M_k + 1)^{( M_k + 1) ( M_{k - 1} + 1) \dotsm ( M_1 + 1) m \xi } \over \Ga (( M_K + 1) \dotsm ( M_1 + 1) m \xi )} \non \\
&\quad \times \{ ( M_K + 1) \dotsm ( M_1 + 1) m \xi \} ^{( M_K + 1) \dotsm ( M_1 + 1) m \xi - m \xi + \sum_{k = 1}^{K} M_k / 2} \non \\
&\quad \times \prod_{k = 1}^{K} \prod_{j = 2}^{M_k + 1} \int_{0}^{\infty } {\tilde{t} _{k, j}}^{( M_{k - 1} + 1) \dotsm ( M_1 + 1) m \xi + (j - 1) / ( M_k + 1) - 1} e^{- \tilde{t} _{k, j} ( M_K + 1) \dotsm ( M_1 + 1) m \xi } d{\tilde{t} _{k, j}} \text{.} \non 
\end{align}
Therefore, 
\begin{align}
{1 \over \{ \Ga ( \xi ) \} ^m} &= {(2 \pi )^{(1 - m) / 2} m^{- 1 / 2} \over \prod_{j = 2}^{m} \Ga ((m - j + 1) / m)} \xi ^{m - 1} m^{m \xi } \Big\{ \prod_{j = 2}^{m} \int_{0}^{1} {\rho _j}^{\xi + (j - 1) / m - 1} (1 - \rho _j )^{(m - j + 1) / m - 1} d{\rho _j} \Big\} \non \\
&\quad \times \Big\{ \prod_{k = 1}^{K} {( M_k + 1)^{- 1 / 2} \over (2 \pi )^{M_k / 2}} \Big\} {e^{( M_K + 1) \dotsm ( M_1 + 1) m \xi } \prod_{k = 1}^{K} ( M_k + 1)^{( M_k + 1) ( M_{k - 1} + 1) \dotsm ( M_1 + 1) m \xi } \over \{ ( M_K + 1) \dotsm ( M_1 + 1) m \xi \} ^{m \xi - \sum_{k = 1}^{K} M_k / 2 - 1 / 2}} \non \\
&\quad \times {\{ ( M_K + 1) \dotsm ( M_1 + 1) m \xi \} ^{( M_K + 1) \dotsm ( M_1 + 1) m \xi - 1 / 2} \over \Ga (( M_K + 1) \dotsm ( M_1 + 1) m \xi ) e^{( M_K + 1) \dotsm ( M_1 + 1) m \xi }} \non \\
&\quad \times \prod_{k = 1}^{K} \prod_{j = 2}^{M_k + 1} \int_{0}^{\infty } {\tilde{t} _{k, j}}^{( M_{k - 1} + 1) \dotsm ( M_1 + 1) m \xi + (j - 1) / ( M_k + 1) - 1} e^{- \tilde{t} _{k, j} ( M_K + 1) \dotsm ( M_1 + 1) m \xi } d{\tilde{t} _{k, j}} \non \\
&= {\prod_{k = 1}^{K} \{ ( M_K + 1) \dotsm ( M_1 + 1) m / (2 \pi ) \} ^{M_k / 2} \over (2 \pi )^{(m - 1) / 2} \prod_{j = 2}^{m} \Ga ((m - j + 1) / m)} {\xi ^{m + \sum_{k = 1}^{K} M_k / 2 + 1 / 2 - 1} e^{( M_K + 1) \dotsm ( M_1 + 1) m \xi } \over \{ ( M_K + 1) \dotsm ( M_1 + 1) \xi \} ^{m \xi }} \non \\
&\quad \times \Big\{ \prod_{k = 1}^{K} ( M_k + 1)^{( M_k + 1) ( M_{k - 1} + 1) \dotsm ( M_1 + 1) m \xi } \Big\} \Big\{ \prod_{j = 2}^{m} \int_{0}^{1} {\rho _j}^{\xi + (j - 1) / m - 1} (1 - \rho _j )^{(m - j + 1) / m - 1} d{\rho _j} \Big\} \non \\
&\quad \times \prod_{k = 1}^{K} \prod_{j = 2}^{M_k + 1} \int_{0}^{\infty } {\tilde{t} _{k, j}}^{( M_{k - 1} + 1) \dotsm ( M_1 + 1) m \xi + (j - 1) / ( M_k + 1) - 1} e^{- \tilde{t} _{k, j} ( M_K + 1) \dotsm ( M_1 + 1) m \xi } d{\tilde{t} _{k, j}} \non \\
&\quad \times {\{ ( M_K + 1) \dotsm ( M_1 + 1) m \xi \} ^{( M_K + 1) \dotsm ( M_1 + 1) m \xi - 1 / 2} \over \Ga (( M_K + 1) \dotsm ( M_1 + 1) m \xi ) e^{( M_K + 1) \dotsm ( M_1 + 1) m \xi }} \text{.} \non 
\end{align}
This proves part (i). 
\hfill$\Box$

\section{Extensions of Theorem 
1}
\label{subsec:extensions}

Parts (i) and (ii) of Theorem \ref{cor:1} follow from parts (i) and (ii) of Theorem \ref{thm:general}, respectively. 
Theorem \ref{cor:1} can be used in a manner similar to that for Theorem 
1. 

\begin{thm}
\label{cor:1} 
Let $m \in \mathbb{N}$ and $K \in \mathbb{N} _0$. 
\begin{itemize}
\item[{\rm{(i)}}]
For all $\xi > 0$, we have 
\begin{align}
{1 \over \{ \Ga ( \xi ) \} ^m} &= C_{m, K} {1 \over \xi ^{m \xi }} \xi ^{m + K / 2 + 1 / 2 - 1} e^{2^K m \xi } \non \\
&\quad \times 2^{\{ 2 ( 2^K - 1) - K \} m \xi } \Big\{ \prod_{j = 2}^{m} \int_{0}^{1} {\rho _j}^{\xi + (j - 1) / m - 1} (1 - \rho _j )^{(m - j + 1) / m - 1} d{\rho _j} \Big\} \non \\
&\quad \times \Big\{ \prod_{k = 1}^{K} \int_{0}^{\infty } {t_k}^{2^{k - 1} m \xi + 1 / 2 - 1} e^{- t_k 2^K m \xi } d{t_k} \Big\} {( 2^K m \xi )^{2^K m \xi - 1 / 2} \over \Ga (2^K m \xi ) e^{2^K m \xi }} \text{,} \non 
\end{align}
where $C_{m, K} = \{ 2^K m / (2 \pi ) \} ^{K / 2} / \big\{ (2 \pi )^{(m - 1) / 2} \prod_{j = 2}^{m} \Ga ((m - j + 1) / m) \big\} $. 
\item[{\rm{(ii)}}]
For all $\xi > 0$, we have 
\begin{align}
{1 \over \{ \Ga ( \xi ) \} ^m} &= C_{m, K} {( \xi + 1)^{K / 2 + 1 / 2 - 1} \over ( \xi + 1)^{m \xi }} \xi ^m e^{2^K m ( \xi + 1)} \non \\
&\quad \times 2^{\{ 2 ( 2^K - 1) - K \} m ( \xi + 1)} \Big\{ \prod_{j = 2}^{m} \int_{0}^{1} {\rho _j}^{\xi + 1 + (j - 1) / m - 1} (1 - \rho _j )^{(m - j + 1) / m - 1} d{\rho _j} \Big\} \non \\
&\quad \times \Big\{ \prod_{k = 1}^{K} \int_{0}^{\infty } {t_k}^{2^{k - 1} m ( \xi + 1) + 1 / 2 - 1} e^{- t_k 2^K m ( \xi + 1)} d{t_k} \Big\} {\{ 2^K m ( \xi + 1) \} ^{2^K m ( \xi + 1) - 1 / 2} \over \Ga ( 2^K m ( \xi + 1)) e^{2^K m ( \xi + 1)}} \text{,} \non 
\end{align}
where $C_{m, K} = \{ 2^K m / (2 \pi ) \} ^{K / 2} / \big\{ (2 \pi )^{(m - 1) / 2} \prod_{j = 2}^{m} \Ga ((m - j + 1) / m) \big\} $. 
\end{itemize}
\end{thm}

As discussed in Section 
2.1, when we use Theorem 
1, we consider $\rho _j$ as additional latent variables, the full conditional distributions of $\rho _j$ are beta,  and we can easily sample $\rho _j$ in a MCMC algorithm. 
When we use Theorem \ref{cor:1}, we consider $\t = ( t_1 , \dots , t_K ) \in (0, \infty )^K$ as a set of latent variables in addition to $\bro $. 
The full conditional distributions of $t_k$ are ${\rm{Ga}} ( t_k | 2^{k - 1} m \xi + 1 / 2, 2^K m \xi )$ if we use part (i) and ${\rm{Ga}} ( t_k | 2^{k - 1} m ( \xi + 1) + 1 / 2, 2^K m ( \xi + 1))$ if we use part (ii). 

Although there are more latent variables, the approximation to the full conditional of $\xi $ becomes more accurate if we use Theorem \ref{cor:1} instead of Theorem 
1. 
For example, in the ${\rm{Ga}} ( \al , \be )$ case considered in Section 
2.2, we have better lower bounds for the acceptance probability. 
That is, we have 
\begin{align}
\exp \Big\{ - {1 \over 12 ( 2^K n \al ^{*} )} \Big\} \ge 1 - {1 \over 12 ( 2^K n \al ^{*} )} \non 
\end{align}
for part (i) and 
\begin{align}
\exp \Big[ - {1 \over 12 \{ 2^K n ( \al ^{*} + 1) \} } \Big] \ge 1 - {1 \over 12 \{ 2^K n ( \al ^{*} + 1) \} } \ge 1 - {1 \over 12 ( 2^K n)} \non 
\end{align}
for part (ii). 
The lower bounds are clearly increasing functions of $K$ and $n$. 
As $K \to \infty $, they converge to $1$ exponentially fast. 
The last bound is independent of all the variables.

When we use part (ii), we have a factor of the form $( \xi + 1)^c$. 
It is rewritten in the following way. 
\begin{itemize}
\item[{\rm{(a)}}]
If $c < 0$, 
\begin{align}
( \xi + 1)^c &= \int_{0}^{\infty } {z^{- c - 1} \over \Ga (- c)} e^{- z ( \xi + 1)} dz \text{.} \non 
\end{align}
\item[{\rm{(b)}}]
If $c = 0$, 
\begin{align}
( \xi + 1)^c = 1 \text{.} \non 
\end{align}
\item[{\rm{(c)}}]
If $c > 0$, 
\begin{align}
( \xi + 1)^c &= \xi ^{c + \ep } \sum_{\ze = 0}^{\infty } \binom{c + \ep + \ze - 1}{\ze } \int_{0}^{\infty } {\eta ^{\ze + \ep - 1} \over \Ga ( \ze + \ep )} e^{- \eta ( \xi + 1)} d\eta \non 
\end{align}
for any $\ep > 0$ by Lemma \ref{lem:4} below. 
\end{itemize}
In case (a), we introduce an additional latent variable $z \in (0, \infty )$. 
Its full conditional distribution is ${\rm{Ga}} (z | - c, \xi + 1)$. 
In case (c), we introduce two latent variables $\ze \in \mathbb{N} _0$ and $\eta \in (0, \infty )$. 
The variable $\eta $ is marginalized out except when we sample $\xi $ and $\eta $ (see van Dyk and Park (2008) and van Dyk and Jiao (2015) for more details). 
The full conditional of $\eta $ is ${\rm{Ga}} ( \eta | \ze + \ep , \xi + 1)$. 
The full conditional of $\ze $ after marginalizing out $\eta $ is proportional to 
\begin{align}
\binom{C + \ep + \ze - 1}{\ze } {1 \over ( \xi + 1)^{\ze + \ep }} &\propto \binom{C + \ep + \ze - 1}{\ze } \Big( {\xi \over \xi + 1} \Big) ^{C + \ep } \Big( {1 \over \xi + 1} \Big) ^{\ze } = {\rm{NB}} \Big( \ze \Big| C + \ep , {\xi \over \xi + 1} \Big) \text{.} \non 
\end{align}

\begin{lem}
\label{lem:4} 
Let $\ep > 0$ and $C > - \ep $. 
Then 
\begin{align}
( \xi _1 + \xi _2 )^C &= {\xi _1}^{C + \ep } \sum_{\ze = 0}^{\infty } \binom{C + \ep + \ze - 1}{\ze } {{\xi _2}^{\ze } \over ( \xi _1 + \xi _2 )^{\ze + \ep }} \non \\
&= {\xi _1}^{C + \ep } \sum_{\ze = 0}^{\infty } \binom{C + \ep + \ze - 1}{\ze } {\xi _2}^{\ze } \int_{0}^{\infty } {\eta ^{\ze + \ep - 1} \over \Ga ( \ze + \ep )} e^{- \eta ( \xi _1 + \xi _2 )} d\eta \non 
\end{align}
for all $\xi > 0$. 
\end{lem}

Finally, we can also use Lemma \ref{lem:normal-IG} in case (c). 

\begin{lem}
\label{lem:normal-IG} 
Let $C > - 1 / 2$. 
Then 
\begin{align}
( \xi _1 + \xi _2 )^C &= \int_{0}^{\infty } {\eta ^{C + 1 / 2 - 1} \over \Ga (C + 1 / 2)} {e^{- \eta / ( \xi _1 + \xi _2 )} \over ( \xi _1 + \xi _2 )^{1 / 2}} d\eta \non \\
&= \int_{0}^{\infty } {\eta ^{C + 1 / 2 - 1} \over \Ga (C + 1 / 2)} \Big[ \int_{- \infty }^{\infty } {1 \over \sqrt{2 \pi }} {\xi _1}^{- 1 / 2} {\xi _2}^{- 1 / 2} e^{- \th ^2 / (2 \xi _1 ) - \{ \th - (2 \eta )^{1 / 2} \} ^2 / (2 \xi _2 )} d\th \Big] d\eta \non \\
&= \int_{0}^{\infty } {\eta ^{C + 1 / 2 - 1} \over \Ga (C + 1 / 2)} \Big\{ \int_{- \infty }^{\infty } {\eta ^{1 / 2} \over \sqrt{\pi }} {\xi _1}^{- 1 / 2} {\xi _2}^{- 1 / 2} e^{- \eta \tilde{\th } ^2 / \xi _1 - \eta ( \tilde{\th } - 1) ^2 / \xi _2} d\tilde{\th } \Big\} d\eta \non 
\end{align}
for all $\xi _1 , \xi _2 > 0$. 
\end{lem}

\section{Proofs of the expressions 
(3), 
(4), (\ref{eq:dir1_prp}), (\ref{eq:nb_prp}), (\ref{eq:wishart_prp_1}), and (\ref{eq:wishart_prp_2})}
\label{subsec:proof_prp} 
Here, we derive the expressions 
(3) and (4) of the main text and the expressions (\ref{eq:dir1_prp}), (\ref{eq:nb_prp}), (\ref{eq:wishart_prp_1}), and (\ref{eq:wishart_prp_2}) of this Supplementary Material. 

\bigskip

\noindent
{\bf Proof of 
(3).} \ \ The posterior of $\th $, $\ta $, and $\al $ is 
\begin{align}
p( \th , \ta , \al | \x ) &\propto p( \al ) {p( \th , \ta ) \over \ta ^{n / 2}} \prod_{i = 1}^{n} \Big[ \al ^{\al } {\Ga ( \al + 1 / 2) \over \Ga ( \al )} / \Big\{ \al + {( x_i - \th )^2 \over 2 \ta } \Big\} ^{\al + 1 / 2} \Big] \non \\
&= {p( \th , \ta ) \over \ta ^{n / 2}} p( \al ) {\al ^{n \al } \over \{ \Ga ( \al ) \} ^n} \prod_{i = 1}^{n} \int_{0}^{\infty } {w_i}^{\al + 1 / 2 - 1} e^{- w_i \{ \al + ( x_i - \th )^2 / (2 \ta ) \} } d{w_i} \text{.} \non 
\end{align}
Then, by Theorem 
1, we have 
\begin{align}
p( \th , \ta , \al | \x ) &\propto {p( \th , \ta ) \over \ta ^{n / 2}} p( \al ) \al ^{n + 1 / 2 - 1} e^{n \al } \Big[ \prod_{i = 1}^{n} \int_{0}^{\infty } {w_i}^{\al + 1 / 2 - 1} e^{- w_i \{ \al + ( x_i - \th )^2 / (2 \ta ) \} } d{w_i} \Big] \non \\
&\quad \times \Big\{ \prod_{i = 2}^{n} \int_{0}^{1} {\rho _i}^{\al + (i - 1) / n - 1} (1 - \rho _i )^{(n - i + 1) / n - 1} d{\rho _i} \Big\} {(n \al )^{n \al - 1 / 2} \over \Ga (n \al ) e^{n \al }} \text{,} \non 
\end{align}
and this completes the proof. 
\hfill$\Box$

\bigskip

\noindent
{\bf Proof of 
(4).} \ \ The posterior of $\p $ and $\bal $ is 
\begin{align}
p( \p , \bal | \x ) &\propto p( \bal ) \Big\{ {\Ga \big ( \sum_{l = 0}^{L} \al _l \big) \over \prod_{l = 0}^{L} \Ga ( \al _l )} \Big\} ^n \prod_{i = 1}^{n} \prod_{l = 0}^{L} {p_{i, l}}^{x_{i, l} + \al _l - 1} \non \\
&= p( \bal ) \Big( \prod_{i = 1}^{n} \prod_{l = 0}^{L} {p_{i, l}}^{x_{i, l} + \al _l - 1} \Big) \Big[ \prod_{l = 0}^{L} {1 \over \{ \Ga ( \al _l ) \} ^n} \Big] \Big( \prod_{i = 1}^{n} \int_{0}^{\infty } {z_i}^{\sum_{l = 0}^{L} \al _l - 1} e^{- z_i} d{z_i} \Big) \text{.} \non 
\end{align}
By Theorem 
1 and Lemma 
1, 
\begin{align}
{1 \over \{ \Ga ( \al _l ) \} ^n} &\propto {\al _l}^n e^{2 n \al _l} \Big\{ \prod_{i = 2}^{n} \int_{0}^{1} {\rho _{i, l}}^{\al _l + (i - 1) / n - 1} (1 - \rho _{i, l} )^{(n - i + 1) / n - 1} d{\rho _{i, l}} \Big\} \non \\
&\quad \times \Big\{ {(n \al _l )^{n \al _l - 1 / 2} \over \Ga (n \al _l ) e^{n \al _l}} \Big\} ^2 \int_{0}^{\infty } {w_l}^{n \al _l - 1} e^{- w_l n {\al _l}^2} d{w_l} \text{.} \non 
\end{align}
Thus, 
\begin{align}
p( \p , \bal | \x ) &\propto p( \bal ) \Big( \prod_{i = 1}^{n} \prod_{l = 0}^{L} {p_{i, l}}^{x_{i, l} + \al _l - 1} \Big) \Big( \prod_{i = 1}^{n} \int_{0}^{\infty } {z_i}^{\sum_{l = 0}^{L} \al _l - 1} e^{- z_i} d{z_i} \Big) \non \\
&\quad \times \prod_{l = 0}^{L} \Big[ {\al _l}^n e^{2 n \al _l} \Big\{ \prod_{i = 2}^{n} \int_{0}^{1} {\rho _{i, l}}^{\al _l + (i - 1) / n - 1} (1 - \rho _{i, l} )^{(n - i + 1) / n - 1} d{\rho _{i, l}} \Big\} \non \\
&\quad \times \Big\{ {(n \al _l )^{n \al _l - 1 / 2} \over \Ga (n \al _l ) e^{n \al _l}} \Big\} ^2 \int_{0}^{\infty } {w_l}^{n \al _l - 1} e^{- w_l n {\al _l}^2} d{w_l} \Big] \text{,} \non 
\end{align}
and the result follows. 
\hfill$\Box$

\bigskip

\noindent
{\bf Proof of (\ref{eq:dir1_prp}).} \ \ The posterior of $\p $ and $\al $ is 
\begin{align}
p( \p , \al | \x ) &\propto p( \al ) \prod _{i = 1}^{n} \Big[ {\Ga ((L + 1) \al ) \over \{ \Ga ( \al ) \} ^{L + 1}} \Big( \prod_{l = 0}^{L} {p_{i, l}}^{\al - 1} \Big) p( \x _i | \p _i ) \Big] \text{.} \label{pdir1p1} 
\end{align}
By Theorem 
1, we have 
\begin{align}
{\Ga ((L + 1) \al ) \over \{ \Ga ( \al ) \} ^{L + 1}} &\propto \Ga ((L + 1) \al ) {\al ^{L + 1 / 2} e^{(L + 1) \al } \over \al ^{(L + 1) \al }} \non \\
&\quad \times \Big\{ \prod_{l = 1}^{L} \int_{0}^{1} {\rho _l}^{\al + l / (L + 1) - 1} (1 - \rho _l )^{(L - l + 1) / (L + 1) - 1} d{\rho _l} \Big\} {\{ (L + 1) \al \} ^{(L + 1) \al - 1 / 2} \over \Ga ((L + 1) \al ) e^{(L + 1) \al }} \non \\
&\propto \al ^{L} (L + 1)^{(L + 1) \al } \prod_{l = 1}^{L} \int_{0}^{1} {\rho _l}^{\al + l / (L + 1) - 1} (1 - \rho _l )^{(L - l + 1) / (L + 1) - 1} d{\rho _l} \text{.} \non 
\end{align}
Therefore, 
\begin{align}
&\prod _{i = 1}^{n} {\Ga ((L + 1) \al ) \over \{ \Ga ( \al ) \} ^{L + 1}} \non \\
&\propto \int _{(0, 1)^{n L}} \al ^{n L} (L + 1)^{n (L + 1) \al } \Big( \prod_{i = 1}^{n} \prod_{l = 1}^{L} {\rho _{i, l}}^{\al } \Big) \Big[ \prod_{i = 1}^{n} \prod_{l = 1}^{L} \{ {\rho _{i, l}}^{l / (L + 1) - 1} (1 - \rho _{i, l} )^{(L - l + 1) / (L + 1) - 1} \} \Big] d\bro \text{.} \non 
\end{align}
Substituting this into (\ref{pdir1p1}) gives 
\begin{align}
p( \p , \al | \x ) &\propto \int _{(0, 1)^{n L}} \Big( p( \al ) \al ^{n L} (L + 1)^{n (L + 1) \al } \Big( \prod_{i = 1}^{n} \prod_{l = 1}^{L} {\rho _{i, l}}^{\al } \Big) \Big( \prod _{i = 1}^{n} \prod_{l = 0}^{L} {p_{i, l}}^{\al } \Big) \non \\
&\quad \times \Big[ \prod_{i = 1}^{n} \prod_{l = 1}^{L} \{ {\rho _{i, l}}^{l / (L + 1) - 1} (1 - \rho _{i, l} )^{(L - l + 1) / (L + 1) - 1} \} \Big] \Big( \prod _{i = 1}^{n} \prod_{l = 0}^{L} {p_{i, l}}^{- 1} \Big) \prod _{i = 1}^{n} p( \x _i | \p _i ) \Big) d\bro \text{,} \non 
\end{align}
where 
\begin{align}
&(L + 1)^{n (L + 1) \al } \Big( \prod_{i = 1}^{n} \prod_{l = 1}^{L} {\rho _{i, l}}^{\al } \Big) \Big( \prod _{i = 1}^{n} \prod_{l = 0}^{L} {p_{i, l}}^{\al } \Big) \non \\
&= \exp \Big( - \al \sum_{i = 1}^{n} \Big[ \sum_{l = 1}^{L} \log {1 \over \rho _{i, l}} + \sum_{l = 0}^{L} \Big\{ \log {1 \over p_{i, l}} - \log (L + 1) \Big\} \Big] \Big) \text{,} \non 
\end{align}
and this completes the proof. 
\hfill$\Box$

\bigskip

\noindent
{\bf Proof of (\ref{eq:nb_prp}).} \ \ The posterior of $\al $ is 
\begin{align}
p( \al | \y ) &\propto p( \al ) {1 \over \{ \Ga ( \al ) \} ^n} \Big\{ \prod_{i = 1}^{n} \Ga ( \al + y_i ) \Big\} \Big( \prod_{i = 1}^{n} p_i \Big) ^{\al } \non \\
&= p( \al ) {1 \over \{ \Ga ( \al ) \} ^n} \Big( \prod_{i = 1}^{n} \int_{0}^{\infty } {z_i}^{\al + y_i - 1} e^{- z_i} d{z_i} \Big) \Big( \prod_{i = 1}^{n} p_i \Big) ^{\al } \text{.} \non 
\end{align}
By Theorem 
1 and Lemma 
1, 
\begin{align}
{1 \over \{ \Ga ( \al ) \} ^n} &\propto \al ^n e^{2 n \al } \Big\{ \prod_{i = 2}^{n} \int_{0}^{1} {\rho _i}^{\al + (i - 1) / n - 1} (1 - \rho _i )^{(n - i + 1) / n - 1} d{\rho _i} \Big\} \non \\
&\quad \times \Big\{ {(n \al )^{n \al - 1 / 2} \over \Ga (n \al ) e^{n \al }} \Big\} ^2 \int_{0}^{\infty } w^{n \al - 1} e^{- w n {\al }^2} dw \text{.} \non 
\end{align}
The desired result follows from the above two expressions. 
\hfill$\Box$

\bigskip

\noindent
{\bf Proof of (\ref{eq:wishart_prp_1}) and (\ref{eq:wishart_prp_2}).} \ \ The posterior of $\bPsi $, $\al $, and $\be $ is 
\begin{align}
p( \bPsi , \al , \be | \x ) &\propto g( \al , \be ) {\al ^{a - 1} e^{- b \al } \be ^{p \{ \al + (p - 1) / 2 \} + c - 1} e^{- d \be } \over 2^{p \al } \Ga _p ( \al + (p - 1) / 2)} | \bPsi |^{n / 2 + \al - 1} e^{- \tr ( \be \I _p + \sum_{i = 1}^{n} \x _i {\x _i}^{\top } ) \bPsi / 2} \text{.} \non 
\end{align}
By making the change of variables $\ga = \be / \al $, we have 
\begin{align}
{p( \bPsi , \al , \ga | \x ) \over g( \al , \al \ga )} &\propto {\al ^{p \{ \al + (p - 1) / 2 \} + c + a - 1} e^{- b \al } \ga ^{p \{ \al + (p - 1) / 2 \} + c - 1} e^{- d \al \ga } \over 2^{p \al } \Ga _p ( \al + (p - 1) / 2)} | \bPsi |^{n / 2 + \al - 1} e^{- \tr ( \al \ga \I _p + \sum_{i = 1}^{n} \x _i {\x _i}^{\top } ) \bPsi / 2} \text{.} \non 
\end{align}
Suppose first that $p = 2 m$ for $m \in \mathbb{N}$. 
Then, by Lemma \ref{lem:wishart} below, 
\begin{align}
{p( \bPsi , \al , \ga | \x ) \over g( \al , \al \ga )} &\propto {\al ^{p \{ \al + (p - 1) / 2 \} + c + a - 1} e^{- b \al } \ga ^{p \{ \al + (p - 1) / 2 \} + c - 1} e^{- d \al \ga } \over 2^{p \al }} | \bPsi |^{n / 2 + \al - 1} e^{- \tr ( \al \ga \I _p + \sum_{i = 1}^{n} \x _i {\x _i}^{\top } ) \bPsi / 2} \non \\
&\quad \times {(2 m)^{2 m \al } \over \Ga (2 m \al )} \prod_{j = 2}^{m} \int_{0}^{1} {\rho _j}^{2 \al + (j - 1) / m - 1} (1 - \rho _j )^{(2 - 1 / m) (j - 1) - 1} d{\rho _j} \non \\
&\propto {e^{2 m \al } \over {\al }^{- 1 / 2}} {\al ^{p (p - 1) / 2 + c + a - 1} e^{- b \al } \ga ^{p \{ \al + (p - 1) / 2 \} + c - 1} e^{- d \al \ga } \over 2^{p \al }} | \bPsi |^{n / 2 + \al - 1} e^{- \tr ( \al \ga \I _p + \sum_{i = 1}^{n} \x _i {\x _i}^{\top } ) \bPsi / 2} \non \\
&\quad \times {(2 m \al )^{2 m \al - 1 / 2} \over \Ga (2 m \al ) e^{2 m \al }} \prod_{j = 2}^{m} \int_{0}^{1} {\rho _j}^{2 \al + (j - 1) / m - 1} (1 - \rho _j )^{(2 - 1 / m) (j - 1) - 1} d{\rho _j} \text{.} \non 
\end{align}
Next, suppose that $p = 2 m - 1$ for $m \in \mathbb{N}$. 
Then, by Lemma \ref{lem:wishart} below, 
\begin{align}
{p( \bPsi , \al , \ga | \x ) \over g( \al , \al \ga )} &\propto {\al ^{p \{ \al + (p - 1) / 2 \} + c + a - 1} e^{- b \al } \ga ^{p \{ \al + (p - 1) / 2 \} + c - 1} e^{- d \al \ga } \over 2^{p \al }} | \bPsi |^{n / 2 + \al - 1} e^{- \tr ( \al \ga \I _p + \sum_{i = 1}^{n} \x _i {\x _i}^{\top } ) \bPsi / 2} \non \\
&\quad \times {(2 m)^{2 m \al } \over \Ga (2 m \al )} \prod_{j = 2}^{m} \int_{0}^{1} {\rho _j}^{2 \al + (j - 1) / m - 1} (1 - \rho _j )^{(2 - 1 / m) (j - 1) - 1} d{\rho _j} \int_{0}^{\infty } z^{\al + m - 1 / 2 - 1} e^{- z} dz \non \\
&\propto {e^{2 m \al } \over {\al }^{1 / 2}} {\al ^{p (p - 1) / 2 + c + a - 1} e^{- b \al } \ga ^{p \{ \al + (p - 1) / 2 \} + c - 1} e^{- d \al \ga } \over 2^{p \al }} | \bPsi |^{n / 2 + \al - 1} e^{- \tr ( \al \ga \I _p + \sum_{i = 1}^{n} \x _i {\x _i}^{\top } ) \bPsi / 2} \non \\
&\quad \times {(2 m \al )^{2 m \al - 1 / 2} \over \Ga (2 m \al ) e^{2 m \al }} \prod_{j = 2}^{m} \int_{0}^{1} {\rho _j}^{2 \al + (j - 1) / m - 1} (1 - \rho _j )^{(2 - 1 / m) (j - 1) - 1} d{\rho _j} \int_{0}^{\infty } z^{\al + m - 1 / 2 - 1} e^{- z} dz \text{.} \non 
\end{align}
This completes the proof. 
\hfill$\Box$

\begin{lem}
\label{lem:wishart} 
Let $p, m \in \mathbb{N}$ and $\al > 0$. 
Then 
\begin{align}
&{\pi ^{p (p - 1) / 4} \over \Ga _p ( \al + (p - 1) / 2)} / {\big\{ \prod_{j = 1}^{m} 2^{2 (j - 1) - 1} \big\} (2 \pi )^{(1 - m) / 2} m^{- 1 / 2} \over \pi ^{m / 2} \prod_{j = 2}^{m} \Ga ((2 - 1 / m) (j - 1))} \non \\
&= \begin{cases} \displaystyle {(2 m)^{2 m \al } \over \Ga (2 m \al )} \prod_{j = 2}^{m} \int_{0}^{1} {\rho _j}^{2 \al + (j - 1) / m - 1} (1 - \rho _j )^{(2 - 1 / m) (j - 1) - 1} d{\rho _j} \text{,} & \text{if $p = 2 m$} \text{,} \\ \displaystyle {(2 m)^{2 m \al } \over \Ga (2 m \al )} \prod_{j = 2}^{m} \int_{0}^{1} {\rho _j}^{2 \al + (j - 1) / m - 1} (1 - \rho _j )^{(2 - 1 / m) (j - 1) - 1} d{\rho _j} \int_{0}^{\infty } z^{\al + m - 1 / 2 - 1} e^{- z} dz \text{,} & \text{if $p = 2 m - 1$} \text{.} \end{cases} \non 
\end{align}
\end{lem}

\noindent
{\bf Proof.} \ \ By the definition of the multivariate gamma function, 
\begin{align}
{\pi ^{p (p - 1) / 4} \over \Ga _p ( \al + (p - 1) / 2)} &= {1 \over \prod_{j = 1}^{p} \Ga ( \al + (p - 1) / 2 - (j - 1) / 2)} \non \\
&= {1 \over \prod_{j = 1}^{p} \Ga ( \al + (p - j) / 2)} = {1 \over \prod_{j = 1}^{p} \Ga ( \al + (j - 1) / 2)} \text{.} \non 
\end{align}
By the duplication formula for the gamma function, 
\begin{align}
\prod_{j = 1}^{2 m} \Ga \Big( \al + {j - 1 \over 2} \Big) &= \prod_{j = 1}^{m} \Big\{ \Ga ( \al + j - 1) \Ga \Big( \al + j - {1 \over 2} \Big) \Big\} = \pi ^{m / 2} \prod_{j = 1}^{m} {\Ga (2 ( \al + j - 1)) \over 2^{2 ( \al + j - 1) - 1}} \text{.} \label{eq:wishart_1} 
\end{align}
Therefore, 
\begin{align}
&{\pi ^{m / 2} / \prod_{j = 1}^{m} 2^{2 ( \al + j - 1) - 1} \over \prod_{j = 1}^{2 m} \Ga ( \al + (j - 1) / 2)} = \prod_{j = 1}^{m} {1 \over \Ga (2 \al + 2 (j - 1))} \non \\
&= {1 \over \Ga (2 \al )} \prod_{j = 2}^{m} \Big\{ {1 \over \Ga (2 \al + 2 (j - 1))} \int_{0}^{1} {\rm{Beta}} \Big( \rho _j \Big| 2 \al + {j - 1 \over m}, \Big( 2 - {1 \over m} \Big) (j - 1) \Big) d{\rho _j} \Big\} \non \\
&= {1 \over \Ga (2 \al )} \prod_{j = 2}^{m} \Big\{ {1 \over \Ga (2 \al + (j - 1) / m) \Ga ((2 - 1 / m) (j - 1))} \int_{0}^{1} {\rho _j}^{2 \al + (j - 1) / m - 1} (1 - \rho _j )^{(2 - 1 / m) (j - 1) - 1} d{\rho _j} \Big\} \text{.} \label{eq:wishart_1.5} 
\end{align}
Then, by Gauss's multiplication formula for the gamma function, it follows that 
\begin{align}
&{\pi ^{m / 2} / \prod_{j = 1}^{m} 2^{2 (j - 1) - 1} \over \prod_{j = 1}^{2 m} \Ga ( \al + (j - 1) / 2)} \prod_{j = 2}^{m} \Ga ((2 - 1 / m) (j - 1)) \non \\
&= {2^{2 m \al } \over \Ga (2 \al )} \prod_{j = 2}^{m} \Big\{ {1 \over \Ga (2 \al + (j - 1) / m)} \int_{0}^{1} {\rho _j}^{2 \al + (j - 1) / m - 1} (1 - \rho _j )^{(2 - 1 / m) (j - 1) - 1} d{\rho _j} \Big\} \non \\
&= 2^{2 m \al } {(2 \pi )^{(1 - m) / 2} m^{2 m \al - 1 / 2} \over \Ga (2 m \al )} \prod_{j = 2}^{m} \int_{0}^{1} {\rho _j}^{2 \al + (j - 1) / m - 1} (1 - \rho _j )^{(2 - 1 / m) (j - 1) - 1} d{\rho _j} \text{.} \label{eq:wishart_2} 
\end{align}
Thus, if $p = 2 m$, 
\begin{align}
&{\pi ^{p (p - 1) / 4} \over \Ga _p ( \al + (p - 1) / 2)} = {1 \over \prod_{j = 1}^{p} \Ga ( \al + (j - 1) / 2)} \non \\
&= {\big\{ \prod_{j = 1}^{m} 2^{2 (j - 1) - 1} \big\} (2 \pi )^{(1 - m) / 2} m^{- 1 / 2} \over \pi ^{m / 2} \prod_{j = 2}^{m} \Ga ((2 - 1 / m) (j - 1))} {(2 m)^{2 m \al } \over \Ga (2 m \al )} \prod_{j = 2}^{m} \int_{0}^{1} {\rho _j}^{2 \al + (j - 1) / m - 1} (1 - \rho _j )^{(2 - 1 / m) (j - 1) - 1} d{\rho _j} \text{.} \non 
\end{align}
On the other hand, if $p = 2 m - 1$, we have, by the duplication formula, 
\begin{align}
&{\pi ^{p (p - 1) / 4} \over \Ga _p ( \al + (p - 1) / 2)} = {1 \over \prod_{j = 1}^{2 m - 1} \Ga ( \al + (j - 1) / 2)} \non \\
&= {1 \over \prod_{j = 1}^{2 m - 2} \Ga ( \al + (j - 1) / 2)} {\Ga ( \al + m - 1 / 2) \over \Ga ( \al + m - 1) \Ga ( \al + m - 1 / 2)} \non \\
&= {1 \over \prod_{j = 1}^{2 (m - 1)} \Ga ( \al + (j - 1) / 2)} {\Ga ( \al + m - 1 / 2) \over \pi ^{1 / 2}} {2^{2 ( \al + m - 1) - 1} \over \Ga (2 ( \al + m - 1))} \non \\
&= {1 \over \pi ^{(m - 1) / 2}} \Big\{ \prod_{j = 1}^{m - 1} {2^{2 ( \al + j - 1) - 1} \over \Ga (2 ( \al + j - 1))} \Big\} {\Ga ( \al + m - 1 / 2) \over \pi ^{1 / 2}} {2^{2 ( \al + m - 1) - 1} \over \Ga (2 ( \al + m - 1))} \non \\
&= {1 \over \pi ^{m / 2}} \Big\{ \prod_{j = 1}^{m} {2^{2 ( \al + j - 1) - 1} \over \Ga (2 ( \al + j - 1))} \Big\} \int_{0}^{\infty } z^{\al + m - 1 / 2 - 1} e^{- z} dz \text{,} \non 
\end{align}
where the fourth equality follows from (\ref{eq:wishart_1}). 
Then, since, by (\ref{eq:wishart_1.5}) and (\ref{eq:wishart_2}), 
\begin{align}
&2^{2 m \al } \prod_{j = 1}^{m} {1 \over \Ga (2 \al + 2 (j - 1))} \prod_{j = 2}^{m} \Ga ((2 - 1 / m) (j - 1)) \non \\
&= 2^{2 m \al } {(2 \pi )^{(1 - m) / 2} m^{2 m \al - 1 / 2} \over \Ga (2 m \al )} \prod_{j = 2}^{m} \int_{0}^{1} {\rho _j}^{2 \al + (j - 1) / m - 1} (1 - \rho _j )^{(2 - 1 / m) (j - 1) - 1} d{\rho _j} \text{,} \non 
\end{align}
we obtain 
\begin{align}
&{\pi ^{p (p - 1) / 4} \over \Ga _p ( \al + (p - 1) / 2)} \non \\
&= {1 \over \pi ^{m / 2}} {\prod_{j = 1}^{m} 2^{2 (j - 1) - 1} \over \prod_{j = 2}^{m} \Ga ((2 - 1 / m) (j - 1))} \int_{0}^{\infty } z^{\al + m - 1 / 2 - 1} e^{- z} dz \non \\
&\quad \times 2^{2 m \al } {(2 \pi )^{(1 - m) / 2} m^{2 m \al - 1 / 2} \over \Ga (2 m \al )} \prod_{j = 2}^{m} \int_{0}^{1} {\rho _j}^{2 \al + (j - 1) / m - 1} (1 - \rho _j )^{(2 - 1 / m) (j - 1) - 1} d{\rho _j} \non \\
&= {\big\{ \prod_{j = 1}^{m} 2^{2 (j - 1) - 1} \big\} (2 \pi )^{(1 - m) / 2} m^{- 1 / 2} \over \pi ^{m / 2} \prod_{j = 2}^{m} \Ga ((2 - 1 / m) (j - 1))} \non \\
&\quad \times {(2 m)^{2 m \al } \over \Ga (2 m \al )} \Big\{ \prod_{j = 2}^{m} \int_{0}^{1} {\rho _j}^{2 \al + (j - 1) / m - 1} (1 - \rho _j )^{(2 - 1 / m) (j - 1) - 1} d{\rho _j} \Big\} \int_{0}^{\infty } z^{\al + m - 1 / 2 - 1} e^{- z} dz \text{.} \non 
\end{align}
This completes the proof. 
\hfill$\Box$

\section{An alternative Metropolis-Hastings algorithm for the model of Section 
3}
\label{subsec:miller_t} 
For the model of Section 
3, we can use the method of Miller (2019) and construct a MH algorithm. 
The posterior of $\th $, $\ta $, and $\al $ is 
\begin{align}
p( \th , \ta , \al | \x ) &\propto p( \al ) {p( \th , \ta ) \over \ta ^{n / 2}} \prod_{i = 1}^{n} \Big[ \al ^{\al } {\Ga ( \al + 1 / 2) \over \Ga ( \al )} / \Big\{ \al + {( x_i - \th )^2 \over 2 \ta } \Big\} ^{\al + 1 / 2} \Big] \non \\
&= {p( \th , \ta ) \over \ta ^{n / 2}} p( \al ) {\al ^{n \al } \over \{ \Ga ( \al ) \} ^n} \prod_{i = 1}^{n} \int_{0}^{\infty } {w_i}^{\al + 1 / 2 - 1} e^{- w_i \{ \al + ( x_i - \th )^2 / (2 \ta ) \} } d{w_i} \text{.} \non 
\end{align}
Then the full conditional distributions of $( \th , \ta )$ and $w_i$ are as given in Section 
3.1. 
The full conditional distribution of $\al $ under the prior $p( \al ) = {\rm{Ga}} ( \al | a, b)$ can be written as 
\begin{align}
{\rm{Ga}} ( \al | a, b) \underbrace{\prod_{i = 1}^{n} {\rm{Ga}} ( w_i | \al , \al )}_{\text{``likelihood"}} \text{.} \non 
\end{align}
Therefore, we can apply the method of Miller (2019) to find numbers $A, B > 0$ such that ${\rm{Ga}} ( \al | A, B)$ is a good approximation to the full conditional density.

\section{Additional results for the simulation studies of Sections 3.2 and 4.2 of the main text}
\label{subsec:additional_figures} 
\begin{figure}[!htbp]
\centering
\includegraphics[width = 16cm]{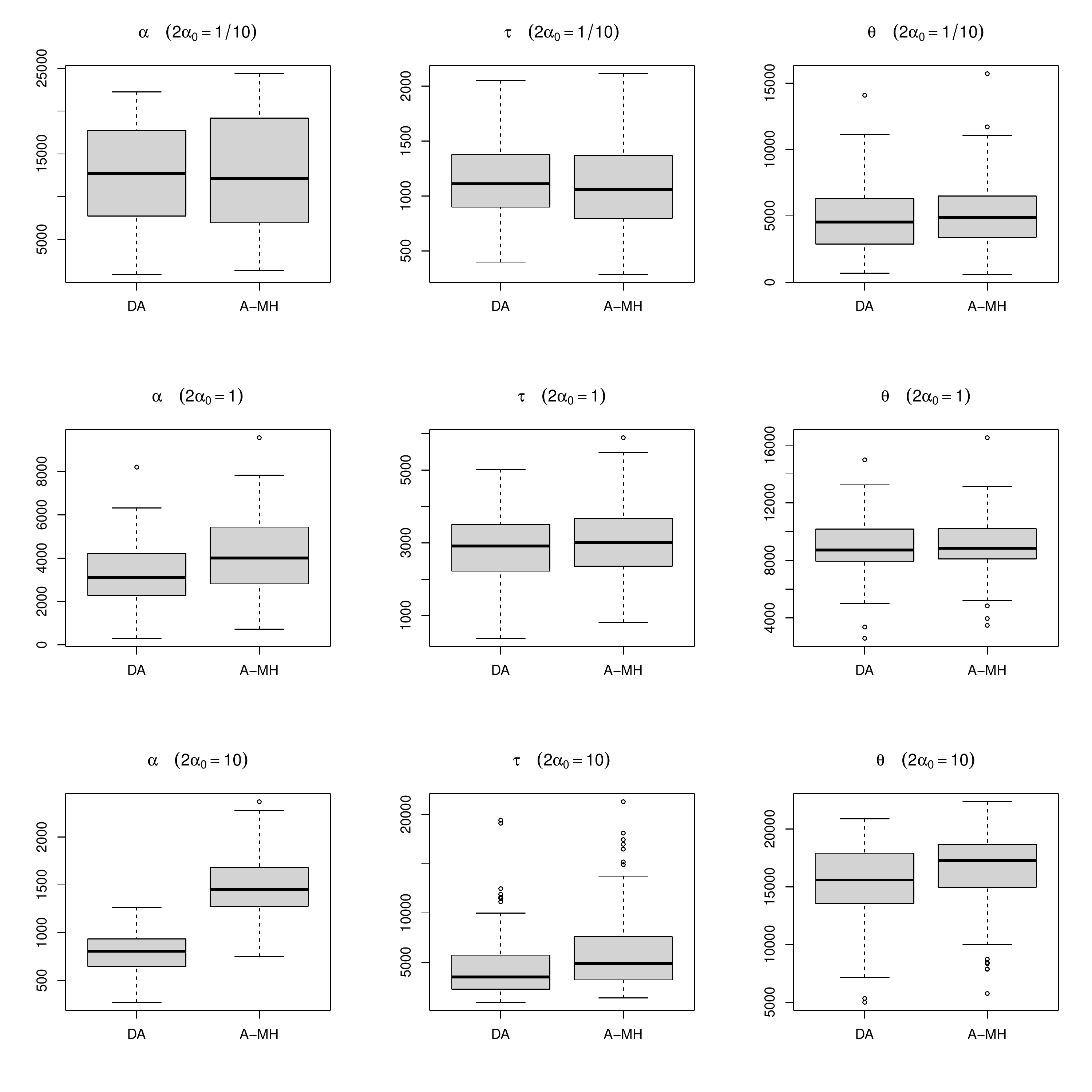}
\caption{Boxplots of the effective sample sizes standardized by the computation times for the proposed method (DA) and the alternative method (A-MH) for $n = 30$. }
\label{fig:t_2}
\end{figure}

\begin{figure}[!htbp]
\centering
\includegraphics[width = 16cm]{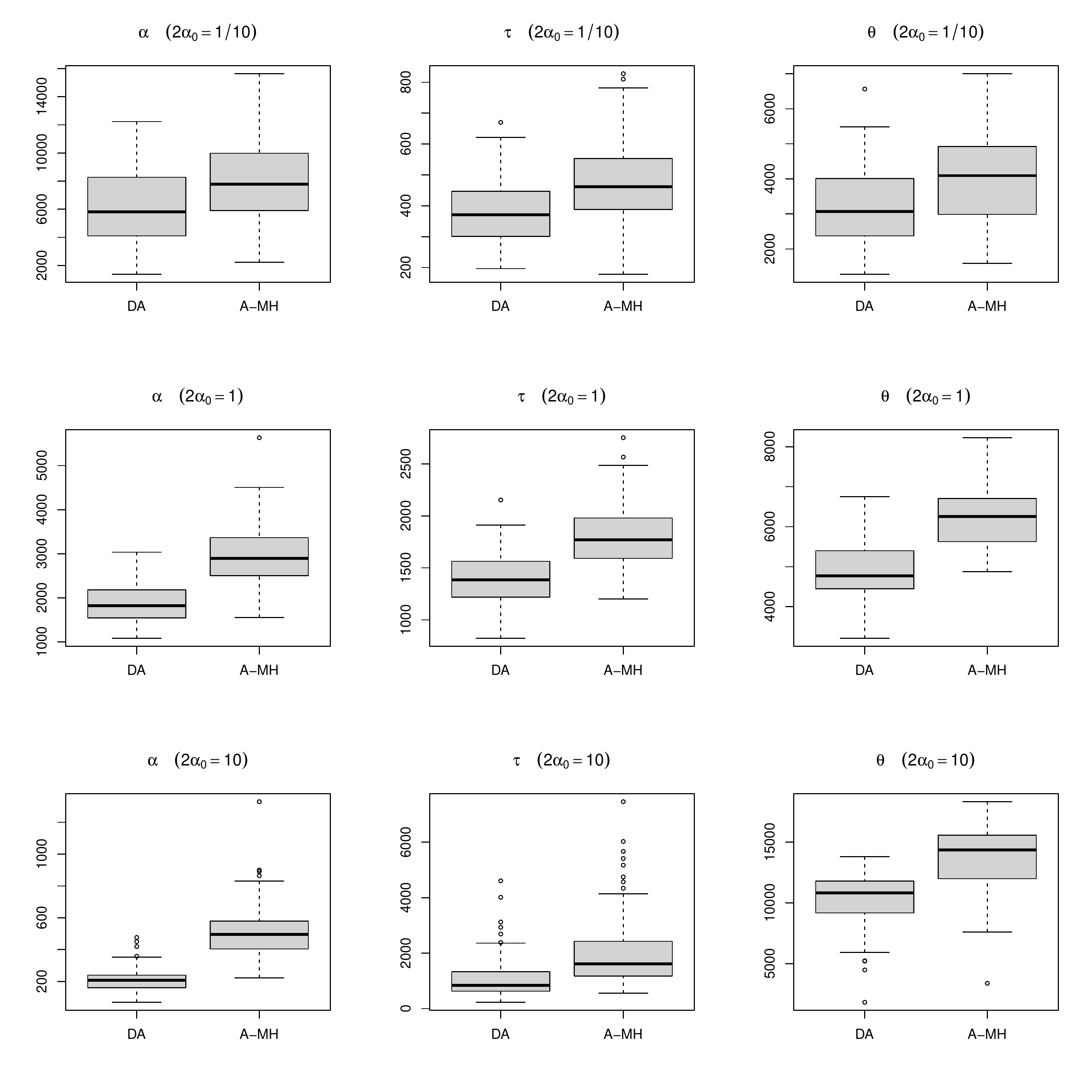}
\caption{Boxplots of the effective sample sizes standardized by the computation times for the proposed method (DA) and the alternative method (A-MH) for $n = 100$. }
\label{fig:t_3}
\end{figure}

\begin{figure}[!htbp]
\centering
\includegraphics[width = 16cm]{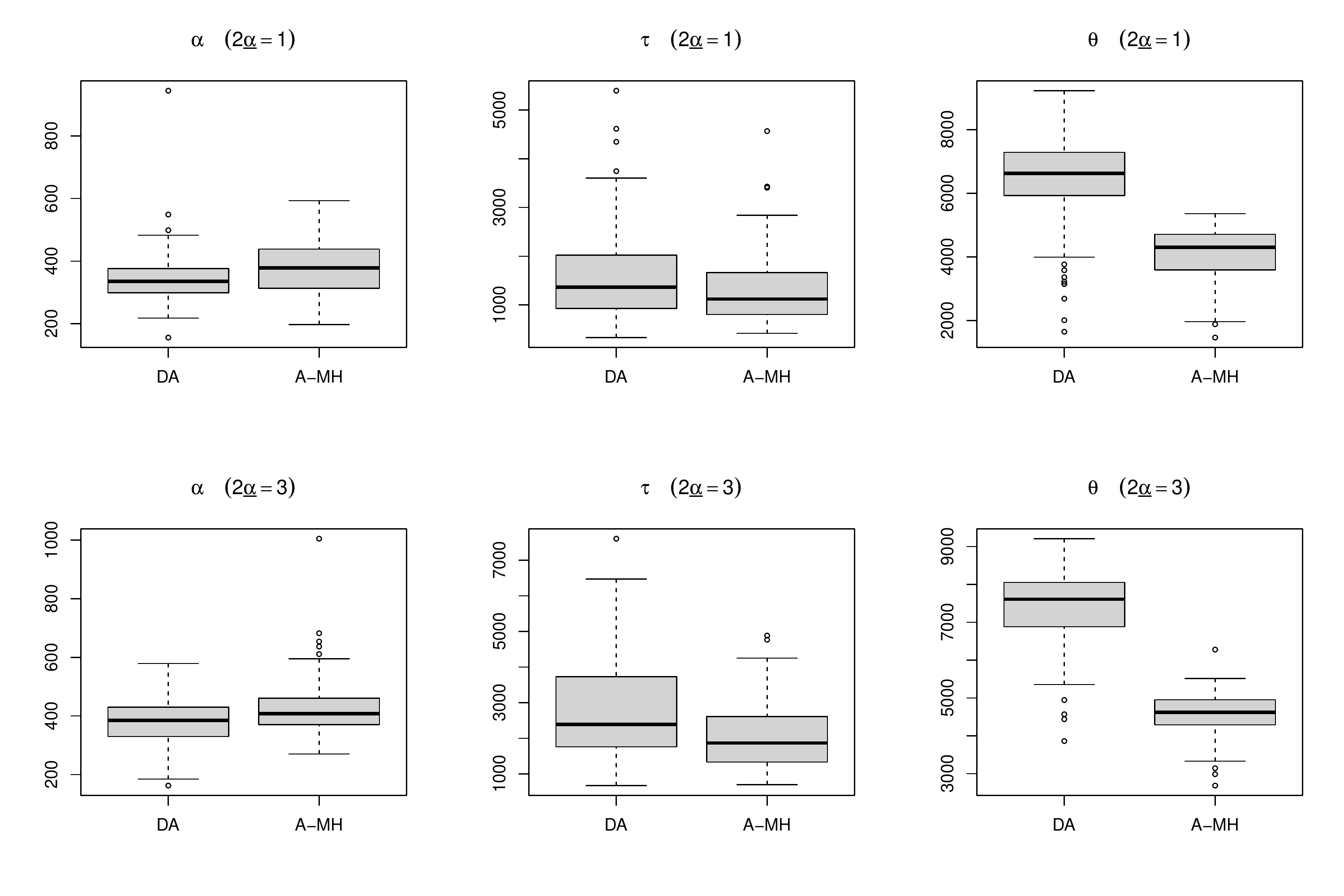}
\caption{Boxplots of the effective sample sizes standardized by the computation times for the proposed method (DA) and the alternative method (A-MH) for $2 \underline{\al } = 1, 3$ for $n = 30$. }
\label{fig:tgamma_2}
\end{figure}

\begin{figure}[!htbp]
\centering
\includegraphics[width = 16cm]{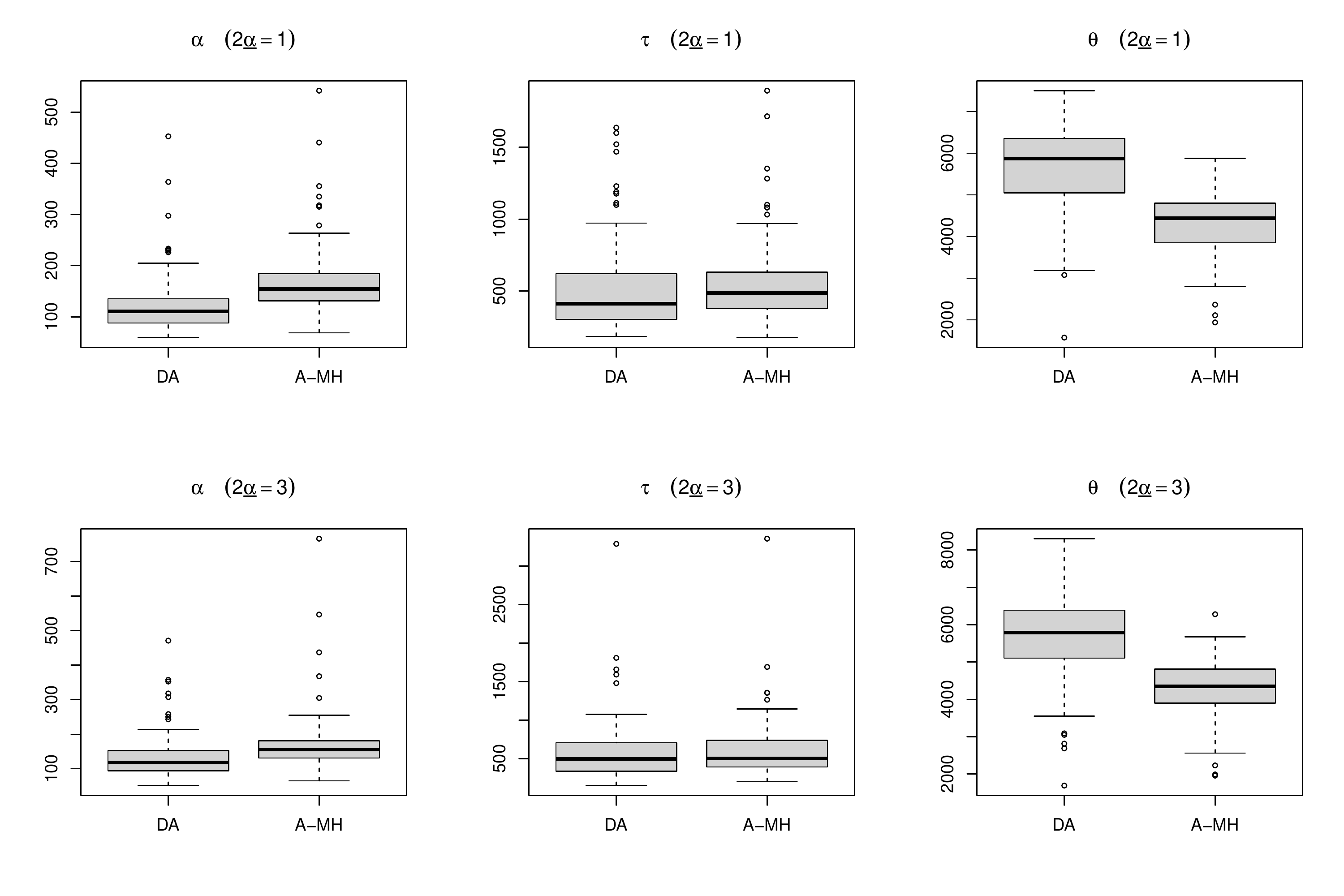}
\caption{Boxplots of the effective sample sizes standardized by the computation times for the proposed method (DA) and the alternative method (A-MH) for $2 \underline{\al } = 1, 3$ for $n = 100$. }
\label{fig:tgamma_3}
\end{figure}

Additional results for the simulation studies of Sections 3.2 and 4.2 of the main text are in this section. 
First, Figures \ref{fig:t_2} and \ref{fig:tgamma_2} correspond to the case of $n = 30$, whereas Figures \ref{fig:t_3} and \ref{fig:tgamma_3} correspond to the case of $n = 100$, and these are as mentioned in the main text. 
Second, Tables S1 and S2 correspond to Table 1 of the main text and show results when we generate $x_i$ from ${\rm{t}} ( x_i | (3, 4), 2 \al _0 )$ and ${\rm{t}} ( x_i | (6, 1), 2 \al _0 )$, whereas Tables S3 and S4 correspond to Table 2 of the main text. 
As in the main text, our method is better in terms of sESS for $\th $ if $n$ is small or if the prior for $\al $ is truncated. 
(Corresponding figures such as Figures 1 and 2 of the main text are omitted because they will not be so different.) 
Finally, Table S5 corresponds to Table 3 of the main text and shows results for two additional scenarios: (III) $\bal _0 = (1, \dots , 1)$ and (IV) $\bal _0 = ((1 / 2, \dots , 1 / 2), (1, \dots , 1)) \in \mathbb{R} ^5 \times \mathbb{R} ^5$. 
It can be seen that our methods show good performance in terms of sESS.

\small
\begin{table}[!thb]
\caption{The averages of the effective sample sizes (ESS) %
for the proposed method (DA) and the alternative method (A-MH) by Miller (2019), the averages of those standardized by computation time (sESS), and the ratios of the mean squared errors (MSE) of the A-MH method to those of the DA method. 
Here, we consider the case of $( \th , \ta ) = (3, 4)$. 
}
\begin{center}
$
{\renewcommand\arraystretch{1.1}\small
\begin{array}{cccccccccccccccccccc}
\hline &&&& \multicolumn{3}{c}{\text{ESS}} && \multicolumn{3}{c}{\text{sESS}} && \multicolumn{3}{c}{\text{MSE ratio}} \\          
n & 2 \al _0 & \text{method} 
&& \text{$\th $} & \text{$\ta $}
& \text{$\al $} 
&& \text{$\th $} & \text{$\ta $}
& \text{$\al $}
&& \text{$\th $} & \text{$\ta $}
& \text{$\al $}
\\

\hline

10 & 	0.1 & \text{DA} &&	832  & 362	 & 1750 	 && 6828  & 	2958 	 & 14213	 && - & -	 & - \\

10 & 	0.1	 & \text{A-MH} &&  802	 & 338& 	1860 && 	6046	 & 2522  & 	13975 	 && 1.04	 & 2.36  & 0.97  \\

10	 & 1 & \text{DA} &&	1312 & 500 	 & 401 && 	11263 & 	4315	 & 3457  && 	- & -	 & - \\

10	 & 1 & \text{A-MH} && 	1314 & 	523  & 	516  && 	10355 	 & 4133  & 	4030 && 0.97 	 & 0.90	 & 1.00  \\

10 & 	10 & \text{DA} &&	1785 & 1053	 & 263  && 	15706	 & 9261 & 	2312  && - & 	-	 & - \\

10	 & 10	 & \text{A-MH} && 1796  & 	1234 & 444  && 	14571  & 	10034	 & 3608 && 	1.00 & 	1.00 	 & 0.98 \\

\hline

30	 & 0.1 & \text{DA} &&	641 	 & 150  & 	1646	 && 3998  & 	936 	 & 10197 && - & - & - \\

30 & 	0.1 & \text{A-MH} && 	645 	 & 148	 & 1628 && 	4121 & 	950 	 & 10427 && 	1.06 	 & 1.19 & 	1.01 \\

30 & 	1 & \text{DA} &&	1256	 & 368 & 	449 && 	7938  & 	2323 & 	2833 && 	- & - & -\\

30	 & 1 & \text{A-MH} && 	1274  & 400 & 	583 	 && 8295 & 2612  & 	3800 && 	0.98 & 	1.00  & 	0.92  \\

30 & 	10 & \text{DA} &&	2122  & 565 & 118 && 	13852 & 	3696  & 768 && 	-& - & -\\

30 & 	10 & \text{A-MH} && 	2226 & 	706 	 & 208	 && 15259 & 	4856	 & 1421	 && 1.01 	 & 1.00  	 & 0.90  \\

\hline

100	 & 0.1	 & \text{DA} && 778  & 	87 & 	1298 && 	2777  & 	310 & 	4622 && - & - & - \\

100	 & 0.1 & \text{A-MH} && 	779	 & 88  & 	1315 	 && 3621 & 	411 	 & 6118	 && 1.00 	 & 1.02	 & 0.99  \\

100	 & 1 & \text{DA} &&	1346  & 	359 & 	489  && 	4922 & 1309 	 & 1784  && 	- & -	 & - \\

100	 & 1 & \text{A-MH} && 	1340	 & 374 & 	609 && 	6311 	 & 1762  & 	2860 	 && 1.00	 & 1.00 	 & 1.01  \\

100 & 	10 & \text{DA} &&	2646 & 	221  & 57&& 	10114& 	849  & 	217  && - & - & 	-\\

100 & 	10 & \text{A-MH} && 	2692  & 337  & 	102  && 13679 	 & 1724	 & 516  && 	1.01 & 	1.00   & 0.80 \\

\hline
\end{array}
}
$
\end{center}
\label{table:S1} 
\end{table}
\normalsize

\small
\begin{table}[!thb]
\caption{The averages of the effective sample sizes (ESS) %
for the proposed method (DA) and the alternative method (A-MH) by Miller (2019), the averages of those standardized by computation time (sESS), and the ratios of the mean squared errors (MSE) of the A-MH method to those of the DA method. 
Here, we consider the case of $( \th , \ta ) = (6, 1)$. 
}
\begin{center}
$
{\renewcommand\arraystretch{1.1}\small
\begin{array}{cccccccccccccccccccc}
\hline &&&& \multicolumn{3}{c}{\text{ESS}} && \multicolumn{3}{c}{\text{sESS}} && \multicolumn{3}{c}{\text{MSE ratio}} \\          
n & 2 \al _0 & \text{method} 
&& \text{$\th $} & \text{$\ta $}
& \text{$\al $} 
&& \text{$\th $} & \text{$\ta $}
& \text{$\al $}
&& \text{$\th $} & \text{$\ta $}
& \text{$\al $}
\\

\hline

10 & 	0.1	 & \text{DA} &&	1000 & 404 	 & 2166	 && 8477  & 	3416& 	18278  && -& 	- & - \\

10	 & 0.1	 & \text{A-MH} &&	894& 	425 	 & 2233 	 && 6922	 & 3268& 	17200 	 && 0.68	 & 35.33 	 & 1.00  \\

10 & 	1 & \text{DA} &&		1823 & 	1010 & 	471 && 	15689  & 8705 	 & 4030 && 	- & - & - \\

10 & 	1 & \text{A-MH} &&		1878& 1081 & 626 && 	14862  & 	8569  & 	4931 	 && 1.01 	 & 1.03 	 & 1.10 \\

10 & 	10 & \text{DA} &&		2745 & 3004 	 & 313  && 	23912	 & 26157 & 	2728  && 	- & - & - \\

10 & 	10 & \text{A-MH} &&		2780 & 	3067 	 & 575	 && 22710 	 & 25109 	 & 4711  && 	1.01  & 1.02 & 1.00  \\

\hline

30 & 	0.1 & \text{DA} &&		1100  & 243 	 & 2542 && 	6856 	 & 1520 & 	15867 && - & 	-	 &-\\

30	 & 0.1 & \text{A-MH} &&		1107 	 & 237& 	2631 	 && 7147  & 1531 & 	16981&& 	0.97  & 0.70 & 1.02  \\

30	 & 1 & \text{DA} &&		1645  & 	577 & 585	 && 10488	 & 3674 & 3727	 && - & - & - \\

30 & 	1 & \text{A-MH} &&		1666& 	591	 & 753	 && 10886 	 & 3848 	 & 4889&& 	1.01& 	1.07  & 0.97  \\

30 & 	10	 & \text{DA} &&	2618  & 1122  & 127  && 17030	 & 7263  & 827&&- & 	-& -\\

30 & 	10	 & \text{A-MH} &&	2674 & 	1310  & 232  && 	18308 	 & 9004 & 	1588  && 	1.01  & 0.98 	 & 0.95  \\

\hline

100 & 	0.1	 & \text{DA} &&	1007  & 140 & 2399	 && 3608  & 	500  & 8585  && 	-& -	 & - \\

100	 & 0.1 & \text{A-MH} &&		995	 & 130 	 & 2448 && 	4621	 & 604& 	11335 && 1.01  & 0.94 	 & 0.99 \\

100 & 	1 & \text{DA} &&		1415 	 & 449 & 	581  && 5227	 & 1652	 & 2132 	 &&- & - & 	- \\

100 & 	1 & \text{A-MH} &&		1427& 	454  & 742  && 6727	 & 2143& 	3493	 && 0.99 	 & 0.99 	 & 0.97\\

100	 & 10	 & \text{DA} &&	2794 & 	349  & 56 	 && 10786 & 	1354	 & 218	 &&-& -& -\\

100	 & 10 & \text{A-MH} &&		2857 & 477 	 & 101  && 	14736 & 	2466	 & 520 	 && 1.01	 & 1.00 	 & 0.91 \\

\hline
\end{array}
}
$
\end{center}
\label{table:S2} 
\end{table}
\normalsize

\small
\begin{table}[!thb]
\caption{The averages of the effective sample sizes (ESS) %
for the proposed method (DA) and the alternative method (A-MH), the averages of those standardized by computation time (sESS), and the ratios of the mean squared errors (MSE) of the A-MH method to those of the DA method for $2 \underline{\al } = 1, 3$. 
Here, we consider the case of $( \th , \ta ) = (3, 4)$. }
\begin{center}
$
{\renewcommand\arraystretch{1.1}\small
\begin{array}{cccccccccccccccccccc}
\hline &&&& \multicolumn{3}{c}{\text{ESS}} && \multicolumn{3}{c}{\text{sESS}} && \multicolumn{3}{c}{\text{MSE ratio}} \\          
n & 2 \underline{\al } & \text{method} 
&& \text{$\th $} & \text{$\ta $}
& \text{$\al $} 
&& \text{$\th $} & \text{$\ta $}
& \text{$\al $}
&& \text{$\th $} & \text{$\ta $}
& \text{$\al $}
\\
\hline

10 & 	1 & \text{DA} &&		1985 	 & 1287	 & 275	 && 6088	 & 3946 & 	842	 && - & 	-	 & - \\

10 & 	1 & \text{A-MH} &&		2046  & 	1445 	 & 467 && 	3432 & 2438	 & 784  && 	0.99 & 	1.00 & 0.90   \\

10 & 	3 & \text{DA} &&		2702  & 	2286 	 & 370	 && 8343 	 & 7052 & 	1142 	 &&- & - & - \\

10	 & 3	 & \text{A-MH} &&	2690  & 	2309 & 	660	 && 4610 	 & 3962	 & 1130 	 && 0.99  & 	1.00  & 	0.94 \\

\hline

30 & 	1	 & \text{DA} &&	2167 & 	505  & 	119 && 	5952	 & 1391  & 	327	 &&-& - & -\\

30 & 	1 & \text{A-MH} &&		2147 & 623 	 & 209	 && 3686 	 & 1077	 & 357  && 	0.99  & 	1.00  & 	0.92  \\

30 & 	3 & \text{DA} &&		2516 & 	979 & 	136	 && 6908	 & 2689 	 & 374&& 	- & - & -\\

30	 & 3 & 	\text{A-MH} &&	2611 	 & 1153 	 & 245	 && 4426 & 	1960  & 415	 && 1.01 	 & 1.00  & 	1.04 \\

\hline

100 & 	1 & \text{DA} &&		2559	 & 211& 	56 && 	5462 & 	450	 & 118 	 && - & -	 & -\\

100	 & 1 & \text{A-MH} &&		2582 & 	316 & 	96	 && 4125 	 & 506 & 	153 && 	0.99 	 & 1.00 	 & 0.94 \\

100	 & 3	 & \text{DA} &&	2729 & 306 & 	59 && 	5803 & 649  & 	124  && 	-& -& -\\

100 & 	3 & \text{A-MH} &&		2752  & 436	 & 106 	 && 4357	 & 692 	 & 168	 && 0.99 & 	1.00 	 & 0.98  \\

\hline
\end{array}
}
$
\end{center}
\label{table:S3} 
\end{table}
\normalsize

\small
\begin{table}[!thb]
\caption{The averages of the effective sample sizes (ESS) %
for the proposed method (DA) and the alternative method (A-MH), the averages of those standardized by computation time (sESS), and the ratios of the mean squared errors (MSE) of the A-MH method to those of the DA method for $2 \underline{\al } = 1, 3$. 
Here, we consider the case of $( \th , \ta ) = (6, 1)$. }
\begin{center}
$
{\renewcommand\arraystretch{1.1}\small
\begin{array}{cccccccccccccccccccc}
\hline &&&& \multicolumn{3}{c}{\text{ESS}} && \multicolumn{3}{c}{\text{sESS}} && \multicolumn{3}{c}{\text{MSE ratio}} \\          
n & 2 \underline{\al } & \text{method} 
&& \text{$\th $} & \text{$\ta $}
& \text{$\al $} 
&& \text{$\th $} & \text{$\ta $}
& \text{$\al $}
&& \text{$\th $} & \text{$\ta $}
& \text{$\al $}
\\

\hline

10 & 	1 & \text{DA} &&		2871  & 3015  & 325  && 	8880 	 & 9324 & 	1004  &&-&- & 	-\\

10	 & 1	 & \text{A-MH} &&	2852  & 	2982	 & 594	 && 4905 & 	5134 & 	1019  && 1.01 & 	0.99 	 & 0.93   \\

10 & 	3 & \text{DA} &&		3121 & 	3228 & 393&& 	9704 & 	10043  & 	1221 	 &&-& -& - \\

10 & 	3 & \text{A-MH} &&		3142  & 	3211 	 & 718 	 && 5379 & 	5500 	 & 1229	 && 1.00  & 	1.00	 & 0.96 \\

\hline

30 & 	1 & \text{DA} &&		2725 & 1272 	 & 130	 && 7561	 & 3531	 & 359 	 && - & - & 	-\\

30 & 	1 & \text{A-MH} &&		2735 & 	1534& 	237	 && 4798 	 & 2696& 	415  && 1.00 & 	1.01  & 	1.06 \\

30 & 	3 & \text{DA} &&		2868  & 	1658	 & 141 && 	8025 	 & 4638	 & 394 && - & 	-	 &-\\

30 & 	3 & 	\text{A-MH} &&	2895 & 1902& 	262 && 	5091 & 	3356 & 460 && 	0.99 & 	1.00  & 	1.00   \\

\hline

100 & 	1 & \text{DA} &&		2808 & 	333  & 	56 	 && 6088& 	724& 	122	 && -& -	 &-\\

100	 & 1 & \text{A-MH} &&		2793  & 437 	 & 101 	 && 4597 	 & 719 	 & 166  && 	1.00 	 & 1.03	 & 0.89   \\

100 & 	3 & \text{DA} &&		2833 & 	336  & 60  && 	6114 & 	726 	 & 130	 && - & -& 	- \\

100 & 	3 & \text{A-MH} &&		2836	 & 502 & 	104&& 4629  & 	823  & 168 && 	1.00	 & 0.99  	 & 1.01 \\

\hline
\end{array}
}
$
\end{center}
\label{table:S4} 
\end{table}
\normalsize

\small
\begin{table}[!thb]
\caption{
The average effective sample size (ESS), the average computation time (CT), the standardized effective sample size by the computation time (sESS), and the mean squared error (MSE) for the proposed data-augmentation method with normal latent variables (DA-N), Poisson latent variables (DA-P), and the PTN sampler of He et al. (2021) (DA-PT) and the original method proposed by He et al. (2021) (ERG). 
These values are averaged over $\al _0 , \dots , \al _9$. 
MSE values under $n=100$ and $n=1000$ are multiplied by $10^3$ and $10^4$, respectively. }
\begin{center}
$
{\renewcommand\arraystretch{1.1}\small
\begin{array}{ccccccccccccccccccc}
\hline
n & \text{Scenario} & \text{method} && \text{ESS} & \text{CT} & \text{sESS} & \text{MSE} \\

\hline

100 & 	\text{(III)} & \text{DA-N} && 	311  & 2.1 & 	151 	 & 10.37 \\

	 &   & 	\text{DA-P} && 448& 	1.7  & 	257  &  10.17  \\

  &   & \text{DA-PT} &&	593 & 	1.8  & 323 	 &  10.19 \\

 	 & & 	\text{ERG} &&	920 & 	82.4  & 	11  &  10.16\\

\hline

100	 & \text{(IV)} & \text{DA-N} && 	564 & 2.1 & 	272& 	 6.40 \\

  &   & \text{DA-P} && 	672& 	1.8 & 381 &  6.51  \\

  &  	 & \text{DA-PT} &&790 & 	1.8 	 & 431  & 	 6.32\\

 	 &   & \text{ERG} &&		1256	 & 82.4 & 15 &  6.35 \\

\hline

1000 & 	\text{(III)}	 &\text{DA-N} && 308& 	7.5	 & 41 	 & 9.65\\

  &  	 & \text{DA-P} && 444	 & 7.1	 & 63	 &  9.66 \\

 &  	 &\text{DA-PT} && 598 	 & 7.1	 & 84 &  9.66 \\

  & 	  &  \text{ERG} &&	918 & 	856.7 & 	1 	 &  9.63  \\

\hline

1000 & 	\text{(IV)} & \text{DA-N} && 	560  & 	7.5  & 	74 & 	6.20  \\

 & 	  & \text{DA-P} && 	670 & 	7.2  & 93  &  6.12 \\

  & 	  & \text{DA-PT} &&	751  & 7.2 & 104  &  6.18\\

 &   & \text{ERG} &&		1231 & 	834.8	 & 1 	 &  6.15   \\

\hline
\end{array}
}
$
\end{center}
\label{table:S5} 
\end{table}
\normalsize

\end{document}